\documentclass{article}

\usepackage[dblblindworkshop,preprint]{neurips_2026}
\usepackage[utf8]{inputenc}
\usepackage[T1]{fontenc}
\usepackage{hyperref}
\usepackage{url}
\usepackage{booktabs}
\usepackage{array}
\usepackage{enumitem}
\usepackage{tabularx}
\usepackage{longtable}
\usepackage{capt-of}
\usepackage{caption}
\usepackage{multirow}
\usepackage{amsmath}
\usepackage{amsfonts}
\usepackage{nicefrac}
\usepackage{microtype}
\usepackage[table]{xcolor}
\usepackage{soul}
\usepackage{graphicx}
\usepackage{flafter}
\usepackage{placeins}
\usepackage{tikz}
\usetikzlibrary{arrows.meta,positioning,calc,fit}
\title{Poison Valleys: Diagnosing Path-Dependent Failure in LLM Code Repair}
\author{Veljko Skarich}
\hypersetup{
  pdftitle={Poison Valleys: Diagnosing Path-Dependent Failure in LLM Code Repair},
  pdfauthor={Veljko Skarich},
  colorlinks=true,
  linkcolor=blue!55!black,
  citecolor=green!40!black,
  urlcolor=blue!55!black
}

\definecolor{traceapi}{HTML}{FFF2CC}
\definecolor{traceorigin}{HTML}{DDEBF7}
\definecolor{tracesurface}{HTML}{E2F0D9}
\definecolor{tracebaseline}{HTML}{FCE4D6}
\definecolor{stagecanonical}{HTML}{4D4D4D}
\definecolor{stagehypothesis}{HTML}{0072B2}
\definecolor{stageplan}{HTML}{E69F00}
\definecolor{stageboth}{HTML}{CC79A7}
\definecolor{semunclassified}{HTML}{999999}
\soulregister\texttt{1}

\begin{document}
\maketitle

\begin{abstract}
When code-repair agents fail, final tests reveal that a patch is wrong but not
which earlier commitment caused it or which software obligation failed. Poison
Valleys traces this path in 1,844 model requests on two synthetic Python cases:
PersistQueue adds atomic batch claiming to a disk-backed queue, while Dijkstra
adds source-sensitive filtering to shortest-path search. We define plausible
but incorrect software mechanisms, place each in a model's hypothesis,
implementation plan, or both, and evaluate the generated code with an Oracle
and targeted diagnostic coordinates. Moving the misconception from hypothesis
to plan increased failure by 32.2\% in PersistQueue and 28.9\% in Dijkstra after
matching the experimental mix. Different misconceptions produced distinct,
model-dependent patterns of violated obligations, although several could
converge on the same behavior. This creates a measured connection from upstream
misconception to code mechanism and failure form. Code, data, and demo:
\href{https://poisonvalleys.org}{\textbf{poisonvalleys.org}}.
\end{abstract}

\section{Introduction}
\label{sec:introduction}

\begin{center}
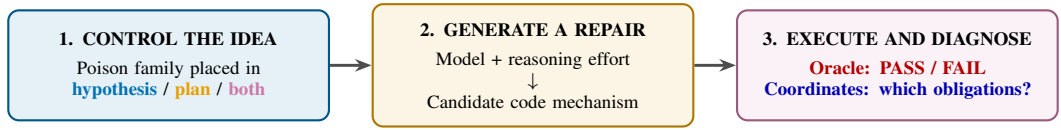

\begin{tikzpicture}[
  font=\scriptsize,
  panel/.style={rounded corners=4pt, minimum height=1.45cm, text width=3.75cm,
    align=center, inner sep=7pt, draw=#1!75!black, fill=#1!10, line width=0.8pt},
  flow/.style={-{Stealth[length=2.2mm]}, line width=1.1pt, draw=stagecanonical}
]
\node[panel=stagehypothesis] (family) {\textbf{1. CONTROL THE IDEA}\\[3pt]
Poison family placed in\\
\textcolor{stagehypothesis}{\textbf{hypothesis}} / \textcolor{stageplan}{\textbf{plan}} / \textcolor{stageboth}{\textbf{both}}};
\node[panel=stageplan, right=0.55cm of family] (model) {\textbf{2. GENERATE A REPAIR}\\[3pt]
Model + reasoning effort\\
$\downarrow$\\
Candidate code mechanism};
\node[panel=stageboth, right=0.55cm of model] (behavior) {\textbf{3. EXECUTE AND DIAGNOSE}\\[3pt]
\textcolor{red!70!black}{\textbf{Oracle: PASS / FAIL}}\\
\textcolor{blue!70!black}{\textbf{Coordinates: which obligations?}}};
\draw[flow] (family) -- (model);
\draw[flow] (model) -- (behavior);
\end{tikzpicture}
\captionof{figure}{From a controlled misconception to executable behavior.}
\label{fig:intro-overview}
\end{center}

A code-repair agent turns a bug report, test output, and repository context into
an explanation of the problem, an implementation plan, and finally a patch. A
passing test can confirm that the resulting program meets the tested contract.
A failing test, however, usually tells us only that the final patch is wrong.
It does not identify which earlier commitment caused the failure, how that
commitment became an implementation choice about state or control flow, or why two
different misconceptions produce different faulty programs. This missing link
limits both scientific explanations of agent behavior and attempts
to repair it.

Prior work can localize failures within agent trajectories, perturb particular
stages, and test whether changing an earlier step improves the outcome
\citep{wang2026star,jaiswal2026obey,ma2026dover,lin2026reflect}. We ask the
forward question: when a particular software misconception enters at a
particular stage, what code mechanism follows, and which executable software
obligations break?

We begin with \emph{poison families}: specific, plausible but incorrect accounts
of how the target software works, together with the implementation commitments
and executable predictions that follow. We define these families before
generation and place one in the model's hypothesis, implementation plan, or
both while holding the repair task fixed. We then trace the path from that
misconception to generated code and executable behavior under each model
configuration. We call these paths \textbf{Poison Valleys} because the supplied
misconception can channel repairs toward a structured region of failures rather
than merely lowering a single success rate.

We construct two synthetic repair cases on pinned versions of real Python
projects. PersistQueue is a disk-backed queue package
\citep{wang2025persistqueue}; our synthetic case adds an atomic batch-claim
operation whose failure must restore durable queue state and remain invisible
to other connections. NetworkX is a graph-analysis library
\citep{hagberg2008networkx}; our synthetic Dijkstra case extends its
multi-source shortest-path implementation with a filter whose decision depends
on the path's original source. Neither case is a recovered historical upstream
bug. Both admit locally plausible repairs that violate a nonlocal obligation.
For each reconstructed candidate, an Oracle determines success or failure.
Independent diagnostic checks then identify which obligations survived, such
as rollback, isolation, record identity, route validity, weight correctness, or
API compatibility. These checks describe the form of a failure without
replacing the Oracle's decision.

The experiment asks three questions. First, does a poison family become more
consequential when it moves from a hypothesis into an implementation plan?
Second, do different poison families produce different patterns of violated
software obligations? Third, does the model and reasoning-effort setting
change how strongly these effects appear? Across \textbf{1,844} model requests,
placing the poison family in the plan increased failure by about
\textbf{30\%} relative to placing it only in the hypothesis in
both tasks. Poison families also produced measurably different distributions of failed obligations, and
the size of the plan effect varied across model configurations.

Poison Valleys maps agent-level concepts to software behavior:
\[
\begin{aligned}
\text{upstream misconception} &\rightarrow \text{implementation commitment}\\
&\rightarrow \text{code mechanism} \rightarrow \text{violated obligation}.
\end{aligned}
\]
The stage identifies where a safeguard can intervene; the violated obligation
identifies what it must protect. A follow-up can return a model's own
failed candidate with a repair prompt naming the violated obligations and test
whether the second candidate passes more checks. Because the original model
responses and downstream executable measurements are preserved, the reported
analysis can be inspected and replayed without making another model request.

\section{Related Work}
\label{sec:related}

Three lines of work frame the experiment: code-repair trajectories, controlled perturbations of agent behavior, and counterfactual failure attribution. They locate errors in an agent trajectory and test whether an intervention changes the outcome. Poison Valleys instead traces a specified misconception at a specified repair stage into generated code and violated software obligations.

\paragraph{Code repair and trajectory analysis.}
Code-repair systems transform issue descriptions, test observations, and repository context into localization decisions, repair reasoning, and patches. SWE-bench evaluates repository-level repairs through executable outcomes \citep{jimenez2024swebench}, while SWE-agent, AutoCodeRover, Agentless, and VRpilot expose distinguishable activities such as repository exploration, fault localization, patch generation, and validation \citep{yang2024sweagent,zhang2024autocoderover,xia2024agentless,kulsum2024vrpilot}. Recent work examines these intermediate activities through failure taxonomies \citep{jiang2026failuremodes}, reconstructed repair rationales \citep{li2026retrace}, dependency-guided repair context \citep{wu2026causalrepair}, and assessments of multi-step behavior beyond terminal reward \citep{ma2024agentboard}. OAT, TRAJDEBUG, AgentRx, and \emph{Failure as a Process} further analyze anomalous actions, error lifecycles, failure onset, and recovery \citep{yeh2026oat,qi2026trajdebug,barke2026agentrx,zhao2026failureprocess}. Together, these studies provide increasingly fine-grained evidence about where a repair trajectory departs from successful behavior. They provide less direct evidence about how a particular upstream software misconception becomes a concrete implementation mechanism and, in turn, determines which executable obligations the resulting program violates. Poison Valleys targets this connection directly.

\paragraph{Stage-wise perturbations and incorrect guidance.}
Controlled interventions reveal how errors propagate through agent trajectories. STAR divides root-cause-analysis traces into stages, locates consequential reasoning faults, and replays stage-specific corrections \citep{wang2026star}. ToolRobustBench aligns perturbations with tool interfaces, user intent, observations, and runtime environments \citep{zheng2026toolrobustbench}. In code repair, \emph{Obey, Diverge, Collapse} finds that incorrect instructions can induce semantic corruption that later self-guided repair does not fix \citep{jaiswal2026obey}.

\paragraph{Counterfactual attribution and executable consequences.}
DoVer, REFLECT, and CausalFlow modify suspected messages, plans, or trajectory steps and replay later behavior \citep{ma2026dover,lin2026reflect,bonagiri2026causalflow}. Project Ariadne intervenes on intermediate reasoning to audit causal and faithfulness claims \citep{khanzadeh2026ariadne}. Those methods start from a failure and repair a suspected earlier step. Poison Valleys runs forward from a specified poison family and repair stage. The Oracle determines correctness; diagnostic checks identify preserved and violated obligations.
\section{Experimental Setup}
\label{sec:experimental-setup}
\label{sec:setup}

\subsection{Experimental design}

The repair sequence is $O \rightarrow D \rightarrow H \rightarrow P \rightarrow C$
(Table~\ref{tab:repair-stages}). Within each case, the task stays fixed while a
reviewed poison family appears in the hypothesis, plan, or both.

\begin{center}
\small
\setlength{\tabcolsep}{5pt}
\begin{tabularx}{\linewidth}{@{}>{\centering\arraybackslash}p{0.07\linewidth}>{\raggedright\arraybackslash}p{0.21\linewidth}X@{}}
\toprule
\textbf{Stage} & \textbf{Name} & \textbf{Definition}\\
\midrule
$\mathbf{O}$ & Observation & The fixed bug report and task requirements shown to the model.\\
$\mathbf{D}$ & Dependency context & Optional information about relevant surrounding code and interfaces.\\
$\mathbf{H}$ & Hypothesis & A proposed explanation of the bug and its underlying software mechanism.\\
$\mathbf{P}$ & Plan & The implementation commitments proposed to repair the bug.\\
$\mathbf{C}$ & Candidate code & The source files returned by the model after receiving the preceding inputs.\\
\bottomrule
\end{tabularx}
\captionof{table}{Stages in the observed repair sequence.}
\label{tab:repair-stages}
\end{center}

A poison family is a specific, plausible but incorrect software idea, such as
assigning transaction state to the wrong owner or losing a route's fixed origin.
Before generation, we align its explanation, implementation commitments,
validating implementation, and executable witness. The validating code is never
shown to the model or used to classify generated code.

\begin{center}
\begin{minipage}{\linewidth}
\small
\setlength{\tabcolsep}{4pt}
\begin{tabularx}{\linewidth}{@{}>{\raggedright\arraybackslash}p{0.16\linewidth}>{\columncolor{stagehypothesis!9}\raggedright\arraybackslash}X>{\columncolor{stageplan!10}\raggedright\arraybackslash}X@{}}
\toprule
& \centering\textcolor{stagehypothesis}{\textbf{HYPOTHESIS ($H$)}} &
\centering\arraybackslash\textcolor{stageplan!80!black}{\textbf{PLAN ($P$)}}\\
\midrule
\textbf{Role} & Explains \emph{why} the bug occurs: a claim about the underlying software mechanism. & Commits to \emph{how} code will change: operations, state ownership, ordering, and error handling.\\
\addlinespace[2pt]
\textbf{PersistQueue A02} & Restoring each claimed row to ready after failure is equivalent to atomic rollback. & Collect staged row IDs; after a SQLite error, write each status back to ready and then re-raise.\\
\addlinespace[2pt]
\textbf{Dijkstra B01} & The shortest-path search found the right winner; the defect exists only in result projection. & Run ordinary multi-source Dijkstra, then reject winning paths that fail the source filter.\\
\bottomrule
\end{tabularx}
\captionof{table}{A hypothesis explains the defect; a plan specifies the code changes.}
\label{tab:hypothesis-plan}
\end{minipage}
\end{center}

\begin{center}
\small
\setlength{\tabcolsep}{4pt}
\begin{tabularx}{\linewidth}{@{}>{\raggedright\arraybackslash}p{0.16\linewidth}>{\raggedright\arraybackslash}p{0.22\linewidth}>{\raggedright\arraybackslash}p{0.22\linewidth}X@{}}
\toprule
\textbf{Condition} & \textbf{Hypothesis ($H$)} & \textbf{Plan ($P$)} & \textbf{What changes}\\
\midrule
\textcolor{stagecanonical}{\rule{0.8em}{0.8em}}~\textbf{Canonical} & Correct explanation & Correct implementation plan & Neither stage is poisoned.\\
\textcolor{stagehypothesis}{\rule{0.8em}{0.8em}}~\textbf{H-only} & \cellcolor{stagehypothesis!10}Poison-family explanation & Correct implementation plan & The wrong account appears only in the explanation.\\
\textcolor{stageplan}{\rule{0.8em}{0.8em}}~\textbf{P-only} & Correct explanation & \cellcolor{stageplan!12}Poison-family commitments & The wrong account directly specifies code choices.\\
\textcolor{stageboth}{\rule{0.8em}{0.8em}}~\textbf{H+P} & \cellcolor{stagehypothesis!10}Poison-family explanation & \cellcolor{stageplan!12}Matching commitments & The same wrong account appears at both stages.\\
\bottomrule
\end{tabularx}
\captionof{table}{Experimental conditions.}
\label{tab:conditions}
\end{center}

Dijkstra also includes no-plan conditions; PersistQueue additionally varies
dependency context. Neither variation defines the main intervention.

Our primary comparison is P-only versus H-only after retaining shared
family--model--dependency combinations and giving both conditions the same mix.
This controls for imbalance across stages. H+P stays separate to show
whether a matching hypothesis further reinforces the plan.

\subsection{Controlled repair tasks}
\label{sec:cases}

Both cases start from real Python packages and add a controlled public feature
plus a faulty implementation. They are not recovered historical
bugs. Table~\ref{tab:case-design} separates what existed upstream from what the
the synthetic extensions add; Appendix~\ref{sec:prompt-traceability} shows the colored task
prompts and their test mappings. PersistQueue is pinned at commit
\texttt{b4fb6d186e375850b2d9ac49fc635779be95f173}; Dijkstra uses NetworkX 3.6.
Both upstream packages are distributed under the BSD 3-Clause license.

\begin{center}
\small
\setlength{\tabcolsep}{4pt}
\begin{tabularx}{\linewidth}{@{}>{\raggedright\arraybackslash}p{0.14\linewidth}>{\raggedright\arraybackslash}p{0.24\linewidth}>{\raggedright\arraybackslash}p{0.27\linewidth}X@{}}
\toprule
\textbf{Case} & \textbf{Existing package} & \textbf{Synthetic extension} & \textbf{Fault and required repair}\\
\midrule
\textbf{PersistQueue} & Disk-backed SQLite queue with single-item \texttt{get}, \texttt{ack}, \texttt{nack}, durable IDs, and FIFO/FILO behavior \citep{wang2025persistqueue}. & \textcolor{stageplan!80!black}{\textbf{\texttt{claim\_batch}}}\newline\texttt{(max\_items, *,}\newline\texttt{raw=False)}, plus a batch-write context and delayed cache publication. & Nested status helpers can commit independently of the apparent outer batch. A correct repair must make selection and every status mutation one rollback unit and publish memory state only after commit.\\
\addlinespace[3pt]
\textbf{Dijkstra} & NetworkX 3.6 multi-source Dijkstra, whose dominance state is keyed by graph node \citep{hagberg2008networkx}. & \textcolor{stagehypothesis}{\textbf{\texttt{source\_filter}}}\newline\texttt{(origin, u, v,}\newline\texttt{edge\_data)} on the public path, path-length, and main APIs. & The faulty search carries \texttt{origin} in the heap but still finalizes by node, discarding a costlier arrival whose origin may permit later edges. A correct repair must preserve origin-specific state until public projection.\\
\bottomrule
\end{tabularx}
\captionof{table}{Existing package, synthetic addition, and repair contract.}
\label{tab:case-design}
\end{center}

\paragraph{PersistQueue.}
Upstream PersistQueue selects and claims one row at a time. The synthetic case adds
\texttt{SQLiteAckQueue.claim\_batch}, a reentrant batch-write context, and a
\texttt{publish} switch that delays updates to the in-memory unacknowledged
cache and count. The faulty method appears to wrap the loop in one
transaction, but each call to the existing decorated status helper can enter
and commit the SQLite connection independently. If the second row fails, the
the first row can remain durably claimed even though the public call
raised. The repair must use one transaction owner for selection and all status
writes, preserve the exact original row IDs and FIFO/FILO order, hide partial
work from other connections, and update cache/count state only after commit.
Appendix~\ref{tab:persistqueue-prompt-tests} maps each part of that contract to
executable tests; Appendix~\ref{tab:pq-family-catalog} gives all eleven plausible
but wrong mechanisms.

\paragraph{Dijkstra.}
Upstream NetworkX 3.6 accepts several sources but has no source-dependent edge
policy. The synthetic case adds the callback
\texttt{source\_filter(origin, u, v, edge\_data)} to three public APIs. Here
\texttt{origin} is the fixed depot where the complete path began; it is not the
currently expanded node. The faulty helper includes the origin in heap
entries but retains ordinary node-keyed \texttt{seen} and \texttt{dist} maps.
That is invalid once future edge permissions depend on origin: a cheaper arrival
from one depot can erase a more expensive arrival from another depot that alone
may legally continue. The repair must keep origin-specific search states, then
project the cheapest permitted result by node while preserving ordinary weights,
cutoffs, exceptions, and no-filter behavior. Appendix~\ref{tab:dijkstra-prompt-tests}
maps the contract to tests; Appendix~\ref{tab:nx-family-catalog} defines all eight
wrong mechanisms.

\subsection{Intervention and model requests}
\label{sec:interventions}

For each request, only the assigned hypothesis and plan change; the task, source
inventory, response format, and evaluation stay fixed. Each request produces one
candidate, so H-only and P-only name where information was supplied, not
separate calls.

The requested model configuration is the model and reasoning-effort
setting sent to the provider. PersistQueue uses GPT-5.6 Luna, GPT-5.4 Nano, and
Claude Haiku 4.5; Dijkstra uses GPT-5.6 Luna, Claude Sonnet 5, and Qwen 3.8
Flash. The primary comparison uses configurations represented in both
conditions. A unique request identifier joins prompt, response, candidate,
execution, and analysis records.

Assignment records what the model received, not what it believed. Generated code
may copy, transform, compensate for, or ignore the supplied mechanism, so we
measure code behavior independently.

\subsection{System design and replay}

\begin{center}
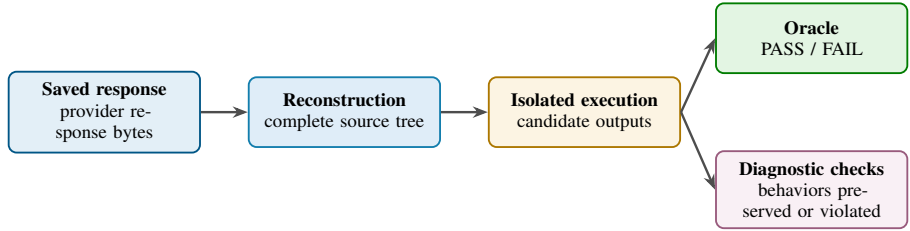

\begin{tikzpicture}[
  font=\scriptsize,
  box/.style={rounded corners=3pt, minimum height=0.92cm, text width=2.25cm,
    align=center, inner sep=4pt, line width=0.7pt, draw=#1!75!black, fill=#1!11},
  arrow/.style={-{Stealth[length=2mm]}, line width=0.9pt, draw=stagecanonical}
]
\node[box=stagehypothesis] (response) {\textbf{Saved response}\\provider response bytes};
\node[box=cyan!70!blue, right=0.62cm of response] (candidate) {\textbf{Reconstruction}\\complete source tree};
\node[box=stageplan, right=0.62cm of candidate] (execute) {\textbf{Isolated execution}\\candidate outputs};
\node[box=green!65!black, above right=0.03cm and 0.45cm of execute] (oracle) {\textbf{Oracle}\\PASS / FAIL};
\node[box=stageboth, below right=0.03cm and 0.45cm of execute] (coordinates) {\textbf{Diagnostic checks}\\behaviors preserved or violated};
\draw[arrow] (response) -- (candidate);
\draw[arrow] (candidate) -- (execute);
\draw[arrow] (execute.east) -- (oracle.west);
\draw[arrow] (execute.east) -- (coordinates.west);
\end{tikzpicture}
\captionof{figure}{Post-generation evaluation and replay.}
\label{fig:evaluation-flow}
\end{center}

Each request UUID joins its prompt, response, source tree, execution, and
analysis row. SHA-256 digests identify the exact records at every handoff.
Replay verifies them, reconstructs the candidate, and reruns execution without
another model call; hashes establish identity, while tests establish behavior.

Reconstruction combines returned files with the expected inventory. Responses
that cannot produce a complete tree remain in request accounting but receive no
generated-code behavior.

\subsection{Oracle and diagnostic checks}
\label{sec:evaluation}

\begin{center}
\small
\setlength{\tabcolsep}{4pt}
\begin{tabularx}{\linewidth}{@{}>{\raggedright\arraybackslash}p{0.17\linewidth}>{\raggedright\arraybackslash}p{0.24\linewidth}>{\raggedright\arraybackslash}p{0.18\linewidth}X@{}}
\toprule
\textbf{Measure} & \textbf{What executes} & \textbf{Output} & \textbf{Example}\\
\midrule
\textcolor{green!50!black}{\textbf{Oracle}} & The complete task protocol compares the candidate's public outputs with expected outputs. & One overall success or failure. & PersistQueue must return the correct batch and leave durable queue state correct after success or failure.\\
\textcolor{stageboth}{\textbf{Coordinate}} & One targeted executable scenario tests one software obligation. & One check passes or fails. & A second connection tries to observe a partially completed claim; hidden state passes the isolation coordinate.\\
\textcolor{stagehypothesis}{\textbf{Axis}} & Related coordinates are grouped by the behavior they test. & An interpretable failure dimension. & Dijkstra shared-junction coordinates contribute to the routing axis because edge legality depends on the original source.\\
\bottomrule
\end{tabularx}
\captionof{table}{Oracle, coordinate, and axis.}
\label{tab:oracle-coordinate-axis}
\end{center}

The \emph{Oracle} determines whether the repair succeeds or fails. For the main
binary outcome, every reconstructed candidate that does
not finish successfully counts as a failure. This includes an ordinary Oracle
failure, a runtime crash, a timeout, or an incomplete evaluation; the specific
reason is recorded separately. Diagnostic checks then test individual software
obligations. PersistQueue uses \textbf{18} selected checks and Dijkstra uses
\textbf{28}. These checks show how a candidate behaved but never overturn the Oracle's correctness
decision. Appendix~\ref{app:diagnostic-catalogs} defines the selected coordinates,
and Table~\ref{tab:failure-subcategories} reports every terminal failure subtype.
When the required checks complete, their results are reduced to the
four task-specific obligation groups in Table~\ref{tab:case-design}. When they
do not complete, we report incomplete diagnostic evidence rather than inventing
a failure pattern.

\begin{center}
\small
\setlength{\tabcolsep}{4pt}
\begin{tabularx}{\linewidth}{@{}lrrrr>{\raggedleft\arraybackslash}X>{\raggedleft\arraybackslash}X@{}}
\toprule
\textbf{Task} & \textbf{Requests} & \textbf{Responses} & \textbf{Candidates} & \textbf{Oracle} & Success & Failure\\
\midrule
PersistQueue & 982 & 843 & 835 & 672 & 453 (\textcolor{green!50!black}{\textbf{54.3\%}}) & 382 (\textcolor{red!70!black}{\textbf{45.7\%}})\\
Dijkstra & 862 & 798 & 798 & 668 & 216 (\textcolor{green!50!black}{\textbf{27.1\%}}) & 582 (\textcolor{red!70!black}{\textbf{72.9\%}})\\
\midrule
\textbf{Total} & 1,844 & 1,641 & 1,633 & 1,340 & 669 (\textcolor{green!50!black}{\textbf{41.0\%}}) & 964 (\textcolor{red!70!black}{\textbf{59.0\%}})\\
\bottomrule
\end{tabularx}
\captionof{table}{Request flow and candidate outcomes.}
\label{tab:request-flow}
\end{center}

Success and failure percentages are calculated from reconstructed candidates;
every reconstructed candidate that does not finish successfully
counts as a failure. Requests without a response or candidate remain visible
in the flow but cannot describe generated-code behavior. The primary comparison
gives H-only and P-only the same mix of represented poison families, model
configurations, and dependency settings. We compute \textbf{95\%} intervals by
resampling complete request UUIDs so that all records from one request remain
together.

All model responses used in the analysis are preserved. Offline replay begins
from a saved response, reconstructs the candidate, and reruns the executable
measurements without making another model request. Code, data, and the interactive
demonstration are available at
\href{https://poisonvalleys.org}{\textbf{poisonvalleys.org}}; detailed family packages,
prompt templates, model settings, executable checks, and analysis tables are
provided in the appendix.

\section{Main Results}
\label{sec:results}

\subsection{Where the poison family appears}

Placing a poison family in the implementation plan caused
substantially more failures than placing it only in the hypothesis. The raw
failure rate rose from \textbf{21.9\%} to \textbf{54.7\%} in PersistQueue and
from \textbf{53.4\%} to \textbf{83.0\%} in Dijkstra
(Table~\ref{tab:primary-stage-comparison}). Because the two conditions did not
contain exactly the same mix of poison families, model
configurations, and dependency settings, we also compared only combinations
represented in both conditions and gave those combinations the same weight.
Under that comparison, plan-only assignment increased failure by
\textbf{32.2\%} in PersistQueue and \textbf{28.9\%} in
Dijkstra. The corresponding failure-risk ratios were
\textbf{2.44 [2.12, 3.23]} and \textbf{1.54 [1.35, 1.88]}: after balancing the
experimental mix, plan-only attempts failed about 2.4 and 1.5 times as often.
The intervals in
Table~\ref{tab:primary-stage-comparison} describe uncertainty across repeated
model requests.

\begin{center}
\scriptsize
\setlength{\tabcolsep}{3pt}
\begin{tabularx}{\linewidth}{@{}l>{\raggedleft\arraybackslash}X>{\raggedleft\arraybackslash}X>{\raggedleft\arraybackslash}p{0.27\linewidth}@{}}
\toprule
\textbf{Task} & \textbf{Hypothesis only} & \textbf{Plan only} & \textbf{Difference with the same experimental mix} \\
\midrule
PersistQueue & \textbf{59/270 (21.9\%)} & \textbf{145/265 (54.7\%)} & \textbf{+32.2\% [27.0\%, 38.8\%]} \\
Dijkstra & \textbf{101/189 (53.4\%)} & \textbf{156/188 (83.0\%)} & \textbf{+28.9\% [21.1\%, 38.6\%]} \\
\bottomrule
\end{tabularx}
\captionof{table}{Failure rates by intervention stage.}
\label{tab:primary-stage-comparison}
\end{center}

Once the poison family was already present in the plan, adding the matching
hypothesis produced a much smaller further change: \textbf{6.3\%} in
PersistQueue \textbf{[0.5\%, 12.2\%]} and \textbf{2.2\%} in Dijkstra
\textbf{[$-$4.9\%, 8.5\%]}. The smaller change is consistent with the plan
acting as the more consequential commitment point;
it does not imply that planning itself is harmful.

\subsection{Which poison family is supplied}

Poison families changed not only how often candidates failed, but also
\emph{which software obligations they violated}. Each diagnostic check tests a
behavior, and related checks are grouped into the eight axes defined
in Table~\ref{tab:diagnostic-axes}. Table~\ref{tab:selected-diagnostic-outcomes}
shows selected PersistQueue plan-only distributions. These checks describe
failure form; they do not
override the Oracle's decision.

The relationship was structured but not deterministic. To summarize its
strength, we use Cram\'er's~$V$, an association score ranging from 0
(no relationship) to 1 (perfect relationship). For plan-only attempts,
the score was \textbf{0.475 [0.453, 0.556]} in PersistQueue and
\textbf{0.314 [0.307, 0.434]} in Dijkstra. Different poison families produced measurably different
distributions of failed obligations, while one family could still produce
several failure patterns and different families could converge on the same
violated obligation.

The fixed-family Dijkstra B01 comparison in
Table~\ref{tab:model-behavior-example} illustrates that model configuration
can also change the outcome.

\begin{center}
\begin{minipage}{\linewidth}
\begingroup
\scriptsize
\setlength{\tabcolsep}{5pt}
\colorbox{stageplan!13}{\parbox{0.94\linewidth}{\centering
\textbf{Experiment key held fixed: Dijkstra $\cdot$ B01 post-filter $\cdot$ plan-only}\quad
Same poison family and stage; only the model configuration differs.}}

\vspace{4pt}
\begin{tabularx}{0.99\linewidth}{@{}>{\raggedright\arraybackslash}p{0.20\linewidth}>{\centering\arraybackslash}p{0.08\linewidth}*{4}{>{\centering\arraybackslash}X}>{\centering\arraybackslash}p{0.15\linewidth}@{}}
\toprule
\textbf{Model} & \textbf{Oracle} & \textcolor{stagehypothesis}{\textbf{Routing}} & \textcolor{stageplan!80!black}{\textbf{Weights}} & \textcolor{stageboth}{\textbf{API}} & \textcolor{green!45!black}{\textbf{Compatibility}} & \textbf{Dijkstra success}\\
\midrule
GPT-5.6 Luna, medium & \colorbox{green!14}{\textcolor{green!45!black}{\textbf{PASS}}} & \textcolor{green!45!black}{Preserved} & \textcolor{green!45!black}{Preserved} & \textcolor{green!45!black}{Preserved} & \textcolor{green!45!black}{Preserved} & 119/340 (\textcolor{green!45!black}{\textbf{35.0\%}})\\
Qwen 3.8 Flash & \colorbox{red!12}{\textcolor{red!70!black}{\textbf{FAIL}}} & \textcolor{red!70!black}{Violated} & \textcolor{red!70!black}{Violated} & \textcolor{green!45!black}{Preserved} & \textcolor{green!45!black}{Preserved} & 6/127 (\textcolor{red!70!black}{\textbf{4.7\%}})\\
\bottomrule
\end{tabularx}
\captionof{table}{The same Dijkstra B01 plan-only intervention produced different behavior across model configurations. The final column reports success across all Dijkstra candidates for each model.}
\label{tab:model-behavior-example}
\endgroup
\end{minipage}
\end{center}

In PersistQueue, a
mechanism that tried to compensate for transaction state in memory was
associated mainly with failures of cross-connection isolation: behavior could
look correct through one connection while another observed the wrong queue
state. In Dijkstra, mechanisms that discarded the fixed route origin while the
search was still running were associated mainly with routing failures at
shared junctions, where the legality of a later edge depends on the path's
origin, linking the supplied misconception to a state decision and an executable
obligation.

We inspected twelve selected generated patches. Five of six
sampled failures and all six sampled passes made a substantive,
mechanism-relevant code change. Because this was a targeted sample,
it illustrates individual pathways and does not estimate how often models adopted
the supplied mechanism across the full corpus.

\subsection{Differences across model configurations}

A model configuration is the model and reasoning-effort setting requested
for an attempt. The plan-only versus hypothesis-only difference varied across
these configurations. In PersistQueue, the differences ranged from
\textbf{$-$3.8\% to +45.1\%}; in Dijkstra, they ranged from
\textbf{+6.6\% to +39.1\%}.
The same intervention can be expressed differently by different
model configurations. These exploratory comparisons do not provide a general
model ranking and do not establish that one model universally reasons better
than another.

The plan-versus-hypothesis result kept the same direction when we
left out one poison family or one model configuration at a time, and when we
restricted the comparison to combinations represented in both conditions that
completed the Oracle. Missing candidates and incomplete diagnostic evidence
remain reported separately rather than being assigned an invented failure
pattern. Detailed condition tables, the model-level success and failure table
(Table~\ref{tab:model-outcomes}), robustness checks, and complete failure
accounting appear in Appendix~\ref{app:visual-placeholders} and
Appendix~\ref{app:statistical-placeholders}.
\section{Discussion}
\label{sec:discussion}

The main result is that a poison family becomes more consequential when
it is expressed as an implementation plan rather than only as an explanation
of the bug. A hypothesis still leaves room to reject the explanation or choose
a safe implementation. A plan commits the repair to operations,
owners of state, update order, error handling, and compatibility choices. When
those commitments are wrong, they can pass directly into code. This does not
mean that planning is harmful. It means that a plan is a high-leverage boundary
where a repair system should check whether the proposed operations preserve the
software contract. The experiment records what information the model received,
not what it believed or internally represented.

Failure rate and failure form answer different questions. The Oracle tells us
whether a candidate satisfies the repair
contract. The diagnostic checks identify the particular obligation that did or
did not survive: for example, restoring queue state after failure, hiding
uncommitted state from another connection, or preserving a route's original
source. Different poison families produced different distributions of
these violated obligations, but the relationship was not one-to-one. One family
could produce several failure patterns, and different families could converge
on the same failed behavior. This is not a classifier that guesses which prompt
input a model saw. It links a specified misconception, through a code mechanism, to an
executable consequence.

\begin{center}
\small
\setlength{\tabcolsep}{6pt}
\begin{tabular}{@{}lrrrrr@{}}
\toprule
\textbf{Family ($n$)} & \textbf{Intact} & \textbf{Isolation only} &
\textbf{Recovery + isolation} & \textbf{Broad} & \textbf{Incomplete}\\
\midrule
A02 ($n=25$) & \textcolor{green!45!black}{\textbf{8\%}} & \cellcolor{stagehypothesis!12}\textcolor{stagehypothesis}{\textbf{92\%}} & 0\% & 0\% & 0\%\\
A08 ($n=25$) & \textcolor{green!45!black}{\textbf{24\%}} & \cellcolor{stagehypothesis!12}\textcolor{stagehypothesis}{\textbf{44\%}} & \cellcolor{stageplan!13}\textcolor{stageplan!80!black}{\textbf{28\%}} & 0\% & \cellcolor{black!8}4\%\\
A14 ($n=24$) & \textcolor{green!45!black}{\textbf{17\%}} & 0\% & \cellcolor{stageplan!13}\textcolor{stageplan!80!black}{\textbf{38\%}} & \cellcolor{red!10}\textcolor{red!70!black}{\textbf{17\%}} & \cellcolor{black!8}29\%\\
\bottomrule
\end{tabular}
\captionof{table}{Diagnostic outcomes for selected PersistQueue poison families in plan-only attempts. The first column gives the code-bearing attempt count $n$; entries are percentages rounded independently.}
\label{tab:selected-diagnostic-outcomes}
\end{center}

In Table~\ref{tab:selected-diagnostic-outcomes}, \emph{intact} means all
selected diagnostic axes passed; \emph{broad} means failures across several
obligation groups; and \emph{incomplete} means the diagnostic panel did not
yield a phenotype. These diagnostic outcomes are separate from the Oracle
verdict.

Generated source can also reveal how a model engaged with a supplied plan. One
A08 plan-only response left the comment, ``\texttt{\_pop} commits each status
update independently,'' and proposed compensating for later claim
failures. That comment is evidence about the mechanism expressed in code, not
a verdict: the Oracle still tests public outputs, and the coordinates still
show which obligations survived.

\begin{center}
\small
\setlength{\tabcolsep}{5pt}
\begin{tabularx}{\linewidth}{@{}>{\raggedright\arraybackslash}p{0.19\linewidth}X@{}}
\toprule
\textbf{Axis} & \textbf{Question}\\
\midrule
\multicolumn{2}{@{}l}{\colorbox{stageplan!12}{\textbf{PersistQueue}}}\\
Recovery & After a database error, are the same rows ready again without an observable intermediate claim?\\
Identity & Does the batch contain the exact selected row IDs, multiplicity, and native order?\\
Isolation & Can another connection observe or interfere with a claim before it commits?\\
Durability & Do database rows, cache/count state, and reopen behavior agree after commit or failure?\\
\midrule
\multicolumn{2}{@{}l}{\colorbox{stagehypothesis!10}{\textbf{Dijkstra}}}\\
Routing & Is every edge legal for the fixed origin that began the path?\\
Weights & Are graph type, edge weights, hidden edges, cycles, and cutoffs handled correctly?\\
API & Do public signatures, return shapes, exceptions, and graph immutability remain correct?\\
Compatibility & Do unfiltered NetworkX behavior and larger branching cases still work?\\
\bottomrule
\end{tabularx}
\captionof{table}{Diagnostic axes by repair task.}
\label{tab:diagnostic-axes}
\end{center}

The mapping depends on the task and model. PersistQueue and Dijkstra showed
the same overall increase when the poison family appeared in the plan, but
they exposed different state, routing, and compatibility obligations. The size
of the increase also varied across model and reasoning-effort settings.
This variation is not a global model ranking. It shows why interventions should
be evaluated against the contract of the software being repaired and against
the model configuration that will use them. A universal label such as
\emph{reasoning failure} would hide the code-level distinction between repairs.

The plan-to-code boundary is a natural intervention point. A repair system can check
whether the plan assigns state to the right owner, preserves rollback and
isolation, carries forward information needed later in execution, and retains
the existing API contract. Whether those checks improve a second repair remains
untested. A follow-up should give the same model its own failed candidate together with a repair
prompt naming the violated obligations, then test whether the second candidate
passes the Oracle and fails fewer or different diagnostic checks.

\FloatBarrier

\section{Limitations}
\label{sec:limitations}

\paragraph{Two controlled repair tasks.}
PersistQueue and Dijkstra were designed to expose specific software obligations
under deterministic execution. That control makes the mechanism-to-code link
inspectable, but it does not represent the full diversity of repositories,
languages, naturally occurring bugs, or misleading information that repair
agents encounter. The repeated direction across the two tasks supports the
experimental method; it does not establish a universal effect size.

\paragraph{Uneven and missing evidence.}
Not every poison family, model configuration, and dependency setting has the
same number of requests, and some combinations are absent. The primary
comparison therefore uses only combinations represented in both the
hypothesis-only and plan-only conditions and gives those combinations the same
weight. This cannot tell us what would happen in combinations we did not run.
Some requests also returned no response or no reconstructable candidate. They
remain in the request accounting and missing-data checks, but they cannot
contribute evidence about generated-code behavior.

\paragraph{Finite executable checks.}
The Oracle and diagnostic checks cover the repair contracts we specified; they
are not complete specifications of PersistQueue or NetworkX. A passing candidate
may contain an untested defect. A failed candidate may stop before every
diagnostic check completes, in which case we report that the failure pattern is
unknown rather than infer one from partial evidence. Visible output and
exception text are even less specific and can make different executable states
look the same.

\paragraph{Assignment, model changes, and uncertainty.}
We know which hypothesis and plan each model received, but we did not
ask human reviewers to decide whether every generated patch believed, copied,
or literally implemented the supplied mechanism. The twelve inspected patches
provide individual examples, not a corpus-wide adoption rate. The model names
also refer to the exact provider configurations available during data
collection; provider updates or routing changes may alter future results.
Repeated requests that share a poison family or experimental
setting are related rather than wholly independent. Our uncertainty intervals
keep records from the same request together, but comparisons between individual
model configurations remain exploratory.

\section{Conclusion}

Poison Valleys connects a specific upstream misconception to the implementation
commitment it creates, the code mechanism that follows, and the executable
obligation it violates. Across \textbf{1,844} model requests on two controlled
repair tasks, placing a poison family in the plan increased failure by
\textbf{32.2\%} in PersistQueue and \textbf{28.9\%} in Dijkstra relative to
placing it only in the hypothesis. Different mechanisms produced different
patterns of violated obligations, and the size of the effect varied across exact
model and reasoning-effort settings.

Inspect a repair after the plan is formed but before it becomes code, and check
the software obligations most at risk. A follow-up should give the same model its own failed candidate plus
a prompt naming the violated obligations, then measure whether the repaired
candidate passes the Oracle and fails fewer diagnostic checks.

\bibliographystyle{plainnat}
\bibliography{references}

@inproceedings{jimenez2024swebench,
  title     = {{SWE-bench}: Can Language Models Resolve Real-World {GitHub} Issues?},
  author    = {Carlos E. Jimenez and John Yang and Alexander Wettig and Shunyu Yao and Kexin Pei and Ofir Press and Karthik Narasimhan},
  booktitle = {International Conference on Learning Representations},
  year      = {2024}
}

@misc{wang2026star,
  title        = {STAR: A Stage-attributed Triage and Repair framework for RCA Agents in Microservices},
  author       = {Junle Wang and Xingchuang Liao and Wenjun Wu},
  year         = {2026},
  eprint       = {2605.15581},
  archivePrefix = {arXiv},
  url          = {https://arxiv.org/abs/2605.15581}
}

@misc{zheng2026toolrobustbench,
  title        = {ToolRobustBench: Stage-Wise Perturbation Evaluation and Failure Diagnosis for Tool-Calling Agents},
  author       = {YiShan Zheng and Yuan Wu and Yi Chang},
  year         = {2026},
  eprint       = {2608.23635},
  archivePrefix = {arXiv},
  url          = {https://arxiv.org/abs/2608.23635}
}

@misc{jaiswal2026obey,
  title        = {Obey, Diverge, Collapse: Blind Obedience to Incorrect Instructions Drives Code LLMs to Irrecoverable Code Semantic Collapse},
  author       = {Raj Jaiswal and Anany Singh Divy and Savar Bhasin and Adi Bajpai and Tanuja Ganu and Rajiv Ratn Shah},
  year         = {2026},
  eprint       = {2607.04537},
  archivePrefix = {arXiv},
  url          = {https://arxiv.org/abs/2607.04537}
}

@misc{lin2026reflect,
  title        = {REFLECT: Intervention-Supported Error Attribution for Silent Failures in LLM Agent Traces},
  author       = {Xiaofeng Lin and Yingxu Wang and Tung Sum Thomas Kwok and Daniel Guo and Sahil Arun Nale and Charles Fleming and Guang Cheng},
  year         = {2026},
  eprint       = {2606.09071},
  archivePrefix = {arXiv},
  primaryClass = {cs.AI},
  url          = {https://arxiv.org/abs/2606.09071}
}

@misc{bonagiri2026causalflow,
  title        = {CausalFlow: Causal Attribution and Counterfactual Repair for LLM Agent Failures},
  author       = {Akash Bonagiri and Devang Borkar and Gerard Janno Anderias and Setareh Rafatirad and Houman Homayoun},
  year         = {2026},
  eprint       = {2605.25338},
  archivePrefix = {arXiv},
  url          = {https://arxiv.org/abs/2605.25338}
}

@misc{khanzadeh2026ariadne,
  title        = {Project Ariadne: A Structural Causal Framework for Auditing Faithfulness in LLM Agents},
  author       = {Sourena Khanzadeh},
  year         = {2026},
  eprint       = {2601.02314},
  archivePrefix = {arXiv},
  url          = {https://arxiv.org/abs/2601.02314}
}

@article{yang2024sweagent,
  title   = {{SWE-agent}: Agent--Computer Interfaces Enable Automated Software Engineering},
  author  = {John Yang and Carlos E. Jimenez and Alexander Wettig and Kilian Lieret and Shunyu Yao and Karthik Narasimhan and Ofir Press},
  journal = {arXiv preprint arXiv:2405.15793},
  year    = {2024}
}

@inproceedings{ma2024agentboard,
  title     = {{AgentBoard}: An Analytical Evaluation Board of Multi-turn {LLM} Agents},
  author    = {Chang Ma and Junlei Zhang and Zhihao Zhu and Cheng Yang and Yujiu Yang and Yaohui Jin and Zhenzhong Lan and Lingpeng Kong and Junxian He},
  booktitle = {Advances in Neural Information Processing Systems},
  volume    = {37},
  year      = {2024}
}

@inproceedings{zhang2024autocoderover,
  title     = {{AutoCodeRover}: Autonomous Program Improvement},
  author    = {Yuntong Zhang and Haifeng Ruan and Zhiyu Fan and Abhik Roychoudhury},
  booktitle = {Proceedings of the 33rd ACM SIGSOFT International Symposium on Software Testing and Analysis},
  pages     = {1592--1604},
  year      = {2024},
  doi       = {10.1145/3650212.3680384}
}

@article{xia2024agentless,
  title   = {Agentless: Demystifying {LLM}-based Software Engineering Agents},
  author  = {Chunqiu Steven Xia and Yinlin Deng and Soren Dunn and Lingming Zhang},
  journal = {arXiv preprint arXiv:2407.01489},
  year    = {2024}
}

@article{kulsum2024vrpilot,
  title   = {A Case Study of {LLM} for Automated Vulnerability Repair: Assessing Impact of Reasoning and Patch Validation Feedback},
  author  = {Ummay Kulsum and Haotian Zhu and Bowen Xu and Marcelo d'Amorim},
  journal = {arXiv preprint arXiv:2405.15690},
  year    = {2024}
}

@article{jiang2026failuremodes,
  title   = {Characterizing the Failure Modes of {LLM}s in Resolving Real-World {GitHub} Issues},
  author  = {Yanjie Jiang and Yian Huang and Guancheng Wang and Junjie Chen and Hui Liu and Lionel Briand},
  journal = {arXiv preprint arXiv:2605.12270},
  year    = {2026}
}

@article{li2026retrace,
  title   = {Independent Patch Verification for Coding Agents with a Bidirectional Reconstruct-and-Verify Framework},
  author  = {Chenglin Li and Yisen Xu and Zehao Wang and Shin Hwei Tan and {Tse-Hsun (Peter) Chen}},
  journal = {arXiv preprint arXiv:2608.08950},
  year    = {2026}
}

@article{wu2026causalrepair,
  title   = {{CausalRepair}: Bridging the Causality Gap in Large Language Model-Based Automated Program Repair via Dual-Slicing},
  author  = {Linhao Wu and Yizhou Chen and Zhen Yang and Pengyu Xue and Dan Hao},
  journal = {arXiv preprint arXiv:2608.10613},
  year    = {2026}
}

@article{yeh2026oat,
  title   = {Tracing Agentic Failure from the Flow of Success},
  author  = {Samuel Yeh and Yiwen Zhu and Shaleen Deep and Sharon Li},
  journal = {arXiv preprint arXiv:2607.12747},
  year    = {2026}
}

@article{qi2026trajdebug,
  title   = {{TRAJDEBUG}: Tracing Error Lifecycle to Identify Critical Failures in Long-Horizon Agent Trajectories},
  author  = {Yunjia Qi and Zehua Yin and Xintong Shi and Hao Peng and Songyuanyi Lu and Yixian Liu and Richeng Xuan and Yuhong Liu and Zhichao Hu and Xiaozhi Wang and Lei Hou and Bin Xu and Juanzi Li},
  journal = {arXiv preprint arXiv:2608.06346},
  year    = {2026}
}

@article{barke2026agentrx,
  title   = {{AgentRx}: Diagnosing {AI} Agent Failures from Execution Trajectories},
  author  = {Shraddha Barke and Arnav Goyal and Alind Khare and Avaljot Singh and Suman Nath and Chetan Bansal},
  journal = {arXiv preprint arXiv:2602.02475},
  year    = {2026},
  url     = {https://arxiv.org/abs/2602.02475}
}

@article{zhao2026failureprocess,
  title   = {Failure as a Process: An Anatomy of {CLI} Coding Agent Trajectories},
  author  = {Xiangxin Zhao and Han Li and Shuaiting Li and Tianyi Zhao and Earl T. Barr and Federica Sarro and He Ye},
  journal = {arXiv preprint arXiv:2607.09510},
  year    = {2026},
  url     = {https://arxiv.org/abs/2607.09510}
}

@inproceedings{ma2026dover,
  title     = {{DoVer}: Intervention-Driven Auto Debugging for {LLM} Multi-Agent Systems},
  author    = {Ming Ma and Jue Zhang and Fangkai Yang and Yu Kang and Qingwei Lin and Saravan Rajmohan and Dongmei Zhang},
  booktitle = {The Fourteenth International Conference on Learning Representations},
  year      = {2026},
  url       = {https://openreview.net/forum?id=mrEK16Jy6h}
}

@misc{wang2025persistqueue,
  title        = {{persist-queue}: A Thread-Safe Disk-Based Persistent Queue in Python},
  author       = {Peter Wang},
  year         = {2025},
  month        = oct,
  howpublished = {\url{https://github.com/peter-wangxu/persist-queue/releases/tag/v1.1.0}},
  note         = {Version 1.1.0, commit b4fb6d1; BSD-3-Clause software}
}

@inproceedings{hagberg2008networkx,
  title     = {Exploring Network Structure, Dynamics, and Function Using {NetworkX}},
  author    = {Aric A. Hagberg and Daniel A. Schult and Pieter J. Swart},
  booktitle = {Proceedings of the 7th Python in Science Conference},
  pages     = {11--15},
  year      = {2008}
}

\appendix
\section{Poison-Family Catalog and Mechanisms}
\label{sec:poison-families}

\subsection{What a poison family specifies}

A poison family is a reviewed, versioned false repair theory, not merely a
short label attached to a prompt. Each family binds four aligned objects: a
plausible false hypothesis about the defect, a plan that operationalizes that
hypothesis, an executable known-wrong realization used during qualification,
and one or more executable witnesses at the public behavioral boundary that
expose the intended mistake. The hypothesis remains diagnosis-only and the plan
remains commitment-only: neither input tells the model that it is poisoned,
names the qualification failure, or explains the contradiction that defeats
it. Exact input bytes and hashes are frozen before use.

The known-wrong realization is a qualification instrument, not a prediction
that generated code will reproduce the same source. It establishes that the
false theory is mechanically coherent, can be implemented without unrelated
syntax or import failure, and is rejected at the intended behavioral boundary.
Under intervention, a model may copy that mechanism, form a hybrid, switch to a
different mechanism, repair the supplied premise, or produce code that cannot
be semantically localized. The empirical object is therefore the complete
attempt-level distribution generated by the prospectively selected family and
stage, not similarity to the known-wrong patch.

A \emph{qualification-witness signature} names the failure pattern that rejects
the known-wrong realization, such as a committed prefix after an exception. It
is not the captured public output, a public-observation vector, or a
\texttt{public\_observation\_hash}. We separately use \emph{family behavioral
signature} for the attempt-level distribution over semantic modes generated by
an intervention condition. The former qualifies the false mechanism; the
latter describes what generated repairs actually do.

Family numbers are identifiers rather than an ordering by severity or
difficulty. Gaps reflect mechanisms that were reserved, merged, rejected, or
not admitted after review. Two families can also share a
qualification-witness signature under the present qualification instrument
while remaining causally distinct interventions. In PersistQueue, A01, A05,
and A07 all admit a
committed-prefix failure, but they attribute it respectively to missing outer
rollback, insufficient preflight, and insufficient locking. In Dijkstra, B03
and B06 both erase fixed-origin identity after the first hop, but one does so
through a first-edge policy and the other through a synthetic-super-source
construction. We retain these families separately because their hypotheses,
implementation commitments, and source mechanisms differ. This separation is
essential for asking which false account a model follows; an output cluster
alone cannot supply that answer.

\subsection{PersistQueue: transactional poison families}

PersistQueue's correct repair must place selection and status mutation inside
one owned transaction, preserve exact row identity and native FIFO/FILO order,
isolate unrelated actors, and publish cache and count state only after the
database commit. The eleven A-families each relax or mislocate a different part
of that contract. Table~\ref{tab:pq-family-catalog} records the model-visible
false account and its induced commitment, followed by the distinguishing
failure for which the executable family was qualified.

\begin{longtable}{@{}p{0.10\textwidth}p{0.42\textwidth}p{0.40\textwidth}@{}}
\toprule
Family & False hypothesis and plan commitment & Defeating behavioral distinction \\
\midrule
\endfirsthead
\toprule
Family & False hypothesis and plan commitment & Defeating behavioral distinction \\
\midrule
\endhead
A01: outer rollback &
The outer batch transaction is assumed to be the only uncommitted unit. The
plan explicitly rolls it back after a database error while leaving the nested
status helpers unchanged. &
Nested helpers can already have committed earlier status writes. Rolling back
the wrapper therefore leaves a claimed prefix visible after the call fails. \\

A02: compensating reset &
Returning every staged row from unacknowledged to ready is treated as equivalent
to rollback because the final status values match. The plan performs reverse
status writes after an error. &
Compensation is itself committed behavior. An external observer can see
ready--unacknowledged--ready transitions, and concurrent actors can interact
with the transiently claimed rows. \\

A03: ownerless transaction &
Existing instance locks are assumed to imply that every helper called while a
queue-level transaction field is populated belongs to the active batch. The
plan routes selection and writes through that field without binding its owner. &
An unrelated same-instance thread can borrow the active transaction, coupling
its write to another caller's commit or rollback; reopen then exposes the
cross-thread state error. \\

A04: read visibility &
Only writes are considered relevant to transaction ownership. The plan routes
decorated writes into the batch transaction but leaves selection on the stale
reader connection because reads are assumed harmless. &
Selection does not see the transaction's own updates. A successful loop can
repeat or lose record identities, and concurrent claimers cease to be
serializable even though the writes share a transaction. \\

A05: preflight validation &
The batch is diagnosed as entering mutation before every row-level status
statement is known to be admissible. The plan preflights selected statements
with SQL \texttt{EXPLAIN} before running the existing per-item loop. &
Static preflight cannot predict a later execution or serialization failure.
Once a real later update fails, earlier per-item commits can remain durable.
This shares a qualified committed-prefix witness signature with A01 and A07. \\

A07: lock-only serialization &
Inconsistent lock lifetime is blamed for partial claims. The plan holds the
instance transaction lock across the entire existing claim loop but preserves
the selection and status-helper implementation. &
Serialization prevents one class of interleaving but does not make individually
committed writes atomic. A single-threaded later failure still leaves a prefix,
and a second queue instance is outside the instance-local lock. \\

A08: reinsert compensation &
Recovery is defined by the multiset of payload values rather than durable record
identity. The plan deletes staged rows and reinserts their payloads as fresh
records after a later error. &
Reinsertion creates replacement IDs and timestamps and can change duplicate and
queue order. Reopen exposes a replacement record rather than restoration of the
original claimed row. \\

A10: wrong-connection savepoint &
The absence of a named rollback checkpoint around the nested calls is treated
as the defect. The plan creates a savepoint on the shared \texttt{\_putter}
connection around the loop and attempts rollback and release during cleanup. &
Nested decorated status helpers commit that same connection, invalidating the
savepoint and separating the apparent savepoint scope from the real atomic
groups. Cleanup can then mask the injected failure with missing-savepoint
state. \\

A11: identity-losing bulk update &
Atomically changing exactly \(N\) claimable rows is assumed sufficient. The plan
bulk-updates rows, then reconstructs the returned IDs by querying the first
\(N\) unacknowledged rows and pairing them with previously selected payloads. &
Preexisting unacknowledged rows, duplicate payloads, smaller batches, and
FIFO/FILO reversal separate mutation count from exact selected identity and
order. \\

A14: retry after failure &
A row-local status error is treated as a transient obstruction rather than a
batch failure. The plan advances past the failing row and continues until the
requested size is reached or no candidate remains. &
The exception is swallowed as a control decision: a committed prefix can
precede the retry, while the returned batch skips the failed identity and
changes native order or size. \\

A16: persistence bypass &
The post-exception inconsistency is attributed to in-memory publication. The
plan snapshots and restores cache and count fields while leaving the existing
database mutation path intact. &
Restoring memory cannot undo durable SQLite status writes. A fresh instance or
reopen observes claimed rows even when the original object's cache and counter
appear restored. \\
\bottomrule
\end{longtable}
\captionof{table}{PersistQueue poison-family mechanisms.}
\label{tab:pq-family-catalog}

The corrected P-only and H+P family counts show that these mechanisms do not
form a simple severity ladder; Table~\ref{tab:selected-diagnostic-outcomes}
gives selected P-only examples. A02 is strongly
isolation-oriented in both stages (23/25 and 23/26). A05 and A10 often combine
recovery and isolation; A11 is also isolation-dominant (21/25 and 19/26). A04
usually lacks a complete coordinate signature (19/24 and 21/24), yet every
terminal failure in those cells still has a specific cause subcategory. This is
a useful counterexample to equating a qualification-witness signature with a
generated-code behavioral signature: the family intervention is known, but the
coordinate instrument need not completely localize every resulting failure.

These distributions clarify what family-level ``resistance'' means. It is an
executable outcome, not a judgment that the model verbally rejected the prompt:
an intact repair may explicitly contradict the supplied plan, silently replace
it with the canonical mechanism, or reach a different correct implementation.
Likewise, an isolation- or broad-dominant family behavioral signature does not
prove literal copying of the qualified patch. The family defines the
intervention; the semantic mode records the downstream behavior.

\subsection{Dijkstra: origin-sensitive search poison families}

In multi-source Dijkstra with an origin-dependent edge policy, a state must
retain every dimension that changes the legality of future transitions. A
source-labeled arrival cannot be discarded solely because another source
reaches the same graph node more cheaply; dominance is valid only within the
state identity used by later filtering. The eight B-families target different
places where this fixed-origin identity can be consulted too late, substituted,
stored only partially, or erased. Table~\ref{tab:nx-family-catalog} gives the
mechanistic catalog.

\begin{longtable}{@{}p{0.10\textwidth}p{0.42\textwidth}p{0.40\textwidth}@{}}
\toprule
Family & False hypothesis and plan commitment & Defeating behavioral distinction \\
\midrule
\endfirsthead
\toprule
Family & False hypothesis and plan commitment & Defeating behavioral distinction \\
\midrule
\endhead
B01: post-filter &
The core search is assumed correct and only its public projection is blamed.
The plan runs ordinary multi-source Dijkstra, reconstructs winners, and removes
paths that fail \texttt{source\_filter} afterward. &
The node-keyed search may already have pruned a more expensive but permitted
source alternative. Deleting the invalid winner cannot reconstruct the valid
route, in either target or all-target output. \\

B02: current-node origin &
The callback's first argument is treated as the node whose adjacency is being
expanded rather than the depot that began the path. The plan passes the current
node as both policy source and edge source. &
The policy then changes meaning at every hop. A single-source reference route
and a shared-junction route distinguish fixed path origin from current search
location. \\

B03: first-edge only &
The source policy is treated as an admission rule for leaving a depot. The plan
checks only the first edge and runs ordinary Dijkstra for the remainder of the
path. &
A later forbidden edge becomes traversable after an allowed first hop. The
source-switching trap exposes this first-hop origin erasure and its effects on
target and all-target routing. \\

B04: partial source state &
Only tentative collisions are blamed. The plan keys provisional distances and
paths by \((origin,node)\) but still finalizes and prunes all alternatives by
node. &
Finalized dominance remains invalid because origin still changes future edge
permission. A later legal continuation can require the source-labeled arrival
discarded at the shared node. \\

B05: paths-only state &
Distances are assumed correct while reconstructed paths alone are said to lose
the depot. The plan stores source-indexed paths but retains one node-level
distance, finalization decision, and heap entry. &
Path identity and distance identity become incoherent: the source used for
filtering or publication need not own the distance that won the node-level
competition. Branching and source-order probes expose the mismatch. \\

B06: super-source &
The search is modeled as a zero-cost synthetic super-source whose branch label
matters only on the first real edge. The plan erases that label afterward and
permits a later edge if any depot would permit it. &
Future permission still depends on the original branch. Its qualification
witness signature is grouped with B03 under first-hop origin erasure, but it
reaches the defect by unioning permissions after a synthetic-source
construction. \\

B12: global edge cache &
Filtering is treated as an expensive, stable property of the directed edge.
The plan runs per-source searches but shares a permission cache keyed only by
\((u,v)\). &
The first depot to inspect an edge determines its visibility for every later
depot. Source-order invariance and false-success probes expose the omitted
origin dimension in the cache key. \\

B13: target fallback &
The combined search is treated as adequate whenever it returns a target. The
plan retries per source only after a target-mode miss, leaving successful target
answers and all-target dictionaries on the original fast path. &
A returned target can still be the wrong permitted route, and an all-target
query never triggers recovery. Target/all-target paired probes expose this
asymmetric projection repair. \\
\bottomrule
\end{longtable}
\captionof{table}{Dijkstra poison-family mechanisms.}
\label{tab:nx-family-catalog}

The corrected Dijkstra family counts keep P-only and H+P separate.
B03 and B13 are the cleanest routing probes:
B03 produces routing phenotypes in 21/25 P-only and 26/26 H+P attempts, while
B13 produces them in 20/23 and 19/23. B02 mixes routing,
weight, intact, and incomplete-panel phenotypes. B04 and B05 likewise show that
partially preserving one component of source state can yield several
downstream forms rather than one family-determined result. Diagnostic
incompleteness remains visible, but it never substitutes for the specific
terminal failure classification.

\subsection{How poison families relate to executable checks}

A poison family names the wrong idea supplied to the model, and the intervention
stage records where that idea was supplied. After the model generates code, the
Oracle answers the binary question: did the repair work? Diagnostic coordinates
are executable checks that answer a finer question: which specific behaviors
survived? We group those checks into four correctness axes---Recovery, Identity,
Isolation, and Durability for PersistQueue; Routing, Weights, API, and
Compatibility for Dijkstra. Each axis asks one question:

\begin{description}[leftmargin=1.9cm,style=nextline]
    \item[Recovery] After something goes wrong, does the queue recover cleanly?
    \item[Identity] Does it claim and return the right records?
    \item[Isolation] Can another actor observe or interfere with partial work?
    \item[Durability] Does the queue remain correct after commit, failure, and reopen?
    \item[Routing] Does it choose a route that is legal for the original source?
    \item[Weights] Does it preserve correct graph and weight behavior?
    \item[API] Does it preserve the public interface and error behavior?
    \item[Compatibility] Does ordinary behavior still work across larger and unfiltered cases?
\end{description}

The first four questions apply to PersistQueue and the second four to Dijkstra.
When the selected panel is complete, the resulting four-axis pattern is the
repair's \emph{executable phenotype}. Incomplete diagnostic evidence has no
phenotype. The Oracle tells us whether the repair worked; the
coordinates show which obligations survived; and only complete coordinate
evidence supports phenotype reduction.

The connection between poison families and coordinates is intentional but not
deterministic. A poisoned plan prescribes a concrete code-level mechanism, whereas
a coordinate executes a behavioral scenario and checks its consequences. The
coordinate evaluator is therefore treatment-label-blind: it receives no
poison-family label and does not inspect the diagnosis or plan for family-specific
language. The shared coordinate panel is nevertheless mechanism-informed.
Several checks were designed to exercise the same task-contract distinctions that
motivated the poison catalog. Thus a coordinate can directly test a consequence
predicted by a family without recognizing that family, and different mechanisms
can fail the same check.

\paragraph{Compensation versus rollback (A02).}
The A02 plan treats returning each staged row from unacknowledged to ready as
equivalent to rollback. The cross-instance failure check instead forces a claim
to fail, pauses the operation while a second queue instance observes and writes
the database, and then checks isolation, rollback, reopen behavior, and later
access. This scenario directly exposes A02's predicted ready--unacknowledged--ready
transition and its interaction with another actor. It does not search the repair
for a compensating-reset loop: A03's shared transaction, A07's instance-local
locking, or another concurrency defect can violate the same executable check.

\paragraph{Payload restoration versus record identity (A08).}
The A08 plan deletes claimed rows and reinserts their payloads after failure,
restoring superficially similar contents under new durable identities. The
\nolinkurl{replacement_compensation} checks seed duplicate payloads, force the
second status update to fail, verify that the original record identities survive,
and verify them again after reopen. These checks are tightly aligned with A08's
predicted consequence but remain implementation-general: any recovery strategy
that recreates values while losing identity, multiplicity, order, or durable
continuity can fail them.

\paragraph{Fixed origin in filtered Dijkstra (B02, B03, and B06).}
B02 substitutes the currently expanded node for the fixed source that began the
route; a single-source reference route distinguishes those callback meanings.
B03 checks only the first edge, while B06 erases synthetic-branch identity and
allows a later edge when any source permits it. The source-switching trap builds a
route that succeeds only if source authority changes or disappears after the
first hop. It can therefore expose both B03 and B06 even though their proposed
code mechanisms differ. These probes test the invariant that every edge is
authorized for one fixed originating source; they do not identify which family,
if any, produced the violation.

This many-to-many relation is the object of measurement rather than a defect in
it. A supplied family may be ignored, partially realized, transformed, or repaired
by the generated program, producing an intact or unexpected complete phenotype.
Conversely, distinct families can converge on the same phenotype because they break
the same correctness axis. RQ2 therefore compares family-associated
distributions over complete executable phenotypes; it does not ask coordinates
to detect a family or infer uptake. The coordinate result is never itself a
poison-family label.

\subsection{What the catalog enables}

The catalog supports three comparisons that aggregate pass rate cannot. First,
the same stage intervention can be compared across false mechanisms: A02 and
A04 both make PersistQueue fail frequently, but one preserves recovery and
identity while breaking isolation, whereas the other produces broad failure.
Second, the same abstract loss can be compared across concrete mechanisms: B03
and B06 share first-hop origin erasure, yet B03 yields a substantially cleaner
routing-dominant behavioral signature. Third, one mechanism can be compared
across models without assuming that global model quality is the explanation.
These are behavioral signatures of controlled conditions, not claims about
hidden internal states.

The catalog also makes negative and incomplete outcomes informative. A high
intact count is consistent with the supplied theory being escaped or repaired
at the executable boundary, without identifying the source-level response. A
mixed signature says that the same prospective condition supports several
downstream outcomes. A high
\texttt{NO\_COMPLETE\_DIAGNOSTIC\_SIGNATURE} count means that the coordinate
instrument cannot always assign a complete phenotype; every terminal failure
in those cells still has a specific cause subcategory.
None of these outcomes licenses redefining a family from the generated source,
an exact vector, a hash, or a post-hoc cluster.
\section{Model-Facing Prompt Traceability}
\label{sec:prompt-traceability}

The following preserved model-facing prompts establish the task obligations
measured by the case-specific Oracles. Highlight colors identify the prompt
clauses represented in the accompanying non-floating traceability tables;
ellipses occur only in abbreviated table excerpts.

\subsection{Dijkstra prompt and tests}
{\scriptsize
\noindent\begin{tabularx}{\textwidth}{|X|}
\hline
NetworkX recently added a {\sethlcolor{traceapi}\hl{\texttt{source\_filter(origin,
u, v, edge\_data)} option}} to \texttt{multi\_source\_dijkstra},
\texttt{multi\_source\_dijkstra\_path}, and
\texttt{multi\_source\_dijkstra\_path\_length} for routing policies that
depend on the depot where a route began. \texttt{origin}
{\sethlcolor{traceorigin}\hl{remains that fixed depot}} for the complete path; a
{\sethlcolor{traceorigin}\hl{truthy result permits the adjacency}} for that origin
and a {\sethlcolor{traceorigin}\hl{falsey result hides it.}} On multigraphs,
{\sethlcolor{tracesurface}\hl{\texttt{edge\_data}}} has the same adjacency-data
shape supplied to a callable weight. The filter must compose with
{\sethlcolor{tracesurface}\hl{string or callable weights}}, including
{\sethlcolor{tracesurface}\hl{hidden edges, cutoff, and target or all-target
results}}; {\sethlcolor{tracesurface}\hl{filter exceptions must propagate
unchanged}} and {\sethlcolor{tracesurface}\hl{calls must not mutate the graph.}}
Combined results must return the {\sethlcolor{traceorigin}\hl{cheapest permitted
route}} across the supplied sources while {\sethlcolor{traceorigin}\hl{retaining
non-exhaustive shortest-path behavior}} on ordinary branching graphs. On a
directed routing graph, a permitted finite route exists from one supplied depot
to a delivery node, but the combined query reports that node as
{\sethlcolor{traceorigin}\hl{unreachable.}} {\sethlcolor{tracebaseline}\hl{Calls that
omit \texttt{source\_filter}}} must retain NetworkX 3.6 behavior. Investigate. \\
\hline
\end{tabularx}\par
}

{\scriptsize
\begin{longtable}{@{}>{\raggedright\arraybackslash}p{0.35\textwidth}>{\raggedright\arraybackslash}p{0.60\textwidth}@{}}
\toprule
Prompt Clause & Tests \\
\midrule
\endfirsthead
\multicolumn{2}{@{}l}{\small Dijkstra prompt-to-test traceability continued}\\
\toprule
Prompt Clause & Tests \\
\midrule
\endhead
\bottomrule
\endfoot
\colorbox{traceapi}{\parbox{\dimexpr\linewidth-2\fboxsep\relax}{\raggedright NetworkX recently added a \texttt{source\_filter}
option to \texttt{multi\_source\_dijkstra}, \ldots\ \texttt{\_path}, and
\texttt{\_path\_length} \ldots}} &
\nolinkurl{single_source_reference_route};
\nolinkurl{two_junction_all_target_projection};
\nolinkurl{graph_and_multigraph_projection};
\nolinkurl{source_identity_or_graph_mutation};
\nolinkurl{medium_branching_projection};
\nolinkurl{callback_exception_propagation} \\
\addlinespace
\colorbox{traceorigin}{\parbox{\dimexpr\linewidth-2\fboxsep\relax}{\raggedright \texttt{origin} remains that fixed depot for the
complete path \ldots}} &
\nolinkurl{shared_junction_combined_distances};
\nolinkurl{shared_junction_combined_paths};
\nolinkurl{shared_junction_source_order_invariance};
\nolinkurl{shared_junction_target_projection}; \nolinkurl{source_switching_trap};
\nolinkurl{two_junction_all_target_projection};
\nolinkurl{two_junction_target_projection};
\nolinkurl{false_success_all_target_projection};
\nolinkurl{false_success_target_projection};
\nolinkurl{late_all_origins_denied} \\
\addlinespace
\colorbox{traceorigin}{\parbox{\dimexpr\linewidth-2\fboxsep\relax}{\raggedright A truthy result permits the adjacency \ldots and a
falsey result hides it.}} & \nolinkurl{source_switching_trap};
\nolinkurl{false_success_all_target_projection};
\nolinkurl{false_success_target_projection};
\nolinkurl{two_junction_all_target_projection};
\nolinkurl{two_junction_target_projection};
\nolinkurl{shared_junction_combined_distances};
\nolinkurl{shared_junction_combined_paths};
\nolinkurl{single_source_reference_route};
\nolinkurl{late_all_origins_denied} \\
\addlinespace
\colorbox{tracesurface}{\parbox{\dimexpr\linewidth-2\fboxsep\relax}{\raggedright On multigraphs, \texttt{edge\_data} has the same
adjacency-data shape supplied to a callable weight.}} &
\nolinkurl{graph_and_multigraph_projection}, specifically
\nolinkurl{_graph_classes} and \nolinkurl{_parallel_edges} \\
\addlinespace
\colorbox{tracesurface}{\parbox{\dimexpr\linewidth-2\fboxsep\relax}{\raggedright The filter must compose with string or callable
weights, including hidden edges \ldots}} &
\nolinkurl{weight_and_cutoff_projection}: missing/default and callable weights,
hidden edges, zero weights, and cycles; filtered negative-weight probe \\
\addlinespace
\colorbox{tracesurface}{\parbox{\dimexpr\linewidth-2\fboxsep\relax}{\raggedright \ldots cutoff \ldots}} &
\nolinkurl{_cutoff_boundaries} within \nolinkurl{weight_and_cutoff_projection} \\
\addlinespace
\colorbox{tracesurface}{\parbox{\dimexpr\linewidth-2\fboxsep\relax}{\raggedright \ldots target or all-target results \ldots}} &
\nolinkurl{shared_junction_target_projection};
\nolinkurl{two_junction_all_target_projection};
\nolinkurl{two_junction_target_projection};
\nolinkurl{false_success_all_target_projection};
\nolinkurl{false_success_target_projection}; path/path-length wrappers; identity
and scaling target calls \\
\addlinespace
\colorbox{tracesurface}{\parbox{\dimexpr\linewidth-2\fboxsep\relax}{\raggedright Filter exceptions must propagate unchanged.}} &
\nolinkurl{callback_exception_propagation}, including exact filter- and
weight-callback exception identity \\
\addlinespace
\colorbox{tracesurface}{\parbox{\dimexpr\linewidth-2\fboxsep\relax}{\raggedright \ldots calls must not mutate the graph.}} &
\nolinkurl{source_identity_or_graph_mutation}; graph-state portions of scaling,
no-filter, validation-error, callback-error, and negative-weight probes \\
\addlinespace
\colorbox{traceorigin}{\parbox{\dimexpr\linewidth-2\fboxsep\relax}{\raggedright Combined results must return the cheapest permitted
route across the supplied sources \ldots}} &
\nolinkurl{competing_route_projection}; shared-junction distance/path checks;
two-junction and false-success checks; \nolinkurl{single_source_reference_route}
\\
\addlinespace
\colorbox{traceorigin}{\parbox{\dimexpr\linewidth-2\fboxsep\relax}{\raggedright \ldots retaining non-exhaustive shortest-path
behavior on ordinary branching graphs.}} & \nolinkurl{medium_branching_projection}
\\
\addlinespace
\colorbox{traceorigin}{\parbox{\dimexpr\linewidth-2\fboxsep\relax}{\raggedright On a directed routing graph, a permitted finite route
exists \ldots but the combined query reports that node as unreachable.}} &
\nolinkurl{single_source_reference_route};
\nolinkurl{shared_junction_combined_distances};
\nolinkurl{shared_junction_combined_paths};
\nolinkurl{shared_junction_target_projection} \\
\addlinespace
\colorbox{tracebaseline}{\parbox{\dimexpr\linewidth-2\fboxsep\relax}{\raggedright Calls that omit \texttt{source\_filter} must
retain NetworkX 3.6 behavior.}} &
\nolinkurl{networkx_3_6_no_filter_regression};
\nolinkurl{networkx_3_6_error_regression}; public validation and unfiltered
negative-weight checks; separate 205-test compatibility panel \\
\addlinespace
Investigate. & No test \\
\end{longtable}
\captionof{table}{Dijkstra prompt-to-test traceability.}
\label{tab:dijkstra-prompt-tests}
}

\subsection{PersistQueue prompt and tests}
{\scriptsize
\noindent\begin{tabularx}{\textwidth}{|X|}
\hline
{\sethlcolor{traceapi}\hl{\texttt{SQLiteAckQueue.claim\_batch(max\_items,
raw=False)} should claim up to the requested number of ready records as one
public operation.}} {\sethlcolor{traceorigin}\hl{If the call raises before
returning a result, no record may have been consumed or left claimed.}}
{\sethlcolor{tracesurface}\hl{A database error during one batch caused the call to
raise without returning records.}} {\sethlcolor{tracesurface}\hl{Afterwards,
\texttt{qsize()} and the persisted ready count disagreed.}}
{\sethlcolor{tracesurface}\hl{Reopening with automatic resumption disabled showed
records that were ready before the call but were unavailable to a later batch
claim.}}
Repair the public behavior while preserving {\sethlcolor{tracebaseline}\hl{FIFO
and FILO order}}, {\sethlcolor{tracebaseline}\hl{duplicate payloads as separate
records}}, {\sethlcolor{tracebaseline}\hl{durable row and status truth across
reopen}}, {\sethlcolor{tracebaseline}\hl{same-database isolation between queue
instances}}, {\sethlcolor{tracebaseline}\hl{safe concurrent claimers}},
{\sethlcolor{tracebaseline}\hl{compatibility with sibling queue types}} and
{\sethlcolor{tracebaseline}\hl{single-item \texttt{get}, \texttt{ack},
\texttt{nack}, and \texttt{ack\_failed} operations}}, and
{\sethlcolor{tracebaseline}\hl{each instance's \texttt{qsize()} under the
baseline contract.}} \\
\hline
\end{tabularx}\par
}

{\scriptsize
\begin{longtable}{@{}>{\raggedright\arraybackslash}p{0.35\textwidth}>{\raggedright\arraybackslash}p{0.60\textwidth}@{}}
\toprule
Prompt Clause & Tests \\
\midrule
\endfirsthead
\multicolumn{2}{@{}l}{\small PersistQueue prompt-to-test traceability continued}\\
\toprule
Prompt Clause & Tests \\
\midrule
\endhead
\bottomrule
\endfoot
\colorbox{traceapi}{\parbox{\dimexpr\linewidth-2\fboxsep\relax}{\raggedright \texttt{SQLiteAckQueue.claim\_batch(max\_items,
raw=False)} should claim up to the requested number of ready records as one
public operation.}} & \nolinkurl{_batch_claim_contract_case};
\nolinkurl{_success_case}; \nolinkurl{_bulk_identity_case} \\
\addlinespace
\colorbox{traceorigin}{\parbox{\dimexpr\linewidth-2\fboxsep\relax}{\raggedright If the call raises before returning a result, no record
may have been consumed or left claimed.}} &
\nolinkurl{_replacement_compensation_case}; \nolinkurl{_retry_after_failure_case};
\nolinkurl{_instance_lock_failure_case}; \nolinkurl{_cross_instance_failure_case};
\nolinkurl{_concurrency_case} \\
\addlinespace
\colorbox{tracesurface}{\parbox{\dimexpr\linewidth-2\fboxsep\relax}{\raggedright A database error during one batch caused the call to
raise without returning records.}} & The same injected-database-failure families:
\nolinkurl{_replacement_compensation_case}; \nolinkurl{_retry_after_failure_case};
\nolinkurl{_instance_lock_failure_case}; \nolinkurl{_cross_instance_failure_case};
\nolinkurl{_concurrency_case} \\
\addlinespace
\colorbox{tracesurface}{\parbox{\dimexpr\linewidth-2\fboxsep\relax}{\raggedright Afterwards, \texttt{qsize()} and the persisted ready
count disagreed.}} & \nolinkurl{_success_case};
\nolinkurl{_concurrent_success_case}; \nolinkurl{_cross_instance_failure_case};
\nolinkurl{_concurrency_case}, especially their persisted-row and public-count
assertions \\
\addlinespace
\colorbox{tracesurface}{\parbox{\dimexpr\linewidth-2\fboxsep\relax}{\raggedright Reopening with automatic resumption disabled showed
records that were ready before the call but were unavailable to a later batch
claim.}} & \nolinkurl{_replacement_compensation_case};
\nolinkurl{_retry_after_failure_case}; \nolinkurl{_success_case};
\nolinkurl{_bulk_identity_case}; \nolinkurl{_cross_instance_failure_case};
\nolinkurl{_concurrency_case} \\
\addlinespace
\colorbox{tracebaseline}{\parbox{\dimexpr\linewidth-2\fboxsep\relax}{\raggedright FIFO and FILO order}} & Every per-queue function;
\nolinkurl{_filo_sqlite_order_case}; order portions of \nolinkurl{_success_case}
and \nolinkurl{_bulk_identity_case} \\
\addlinespace
\colorbox{tracebaseline}{\parbox{\dimexpr\linewidth-2\fboxsep\relax}{\raggedright duplicate payloads as separate records}} &
\nolinkurl{_replacement_compensation_case}; \nolinkurl{_retry_after_failure_case};
\nolinkurl{_success_case}; \nolinkurl{_bulk_identity_case} \\
\addlinespace
\colorbox{tracebaseline}{\parbox{\dimexpr\linewidth-2\fboxsep\relax}{\raggedright durable row and status truth across reopen}} & All
reopen portions of the atomic and concurrency functions;
\nolinkurl{_sqlite_update_persistence_case}; \nolinkurl{_sqlite_delete_reopen_case}
\\
\addlinespace
\colorbox{tracebaseline}{\parbox{\dimexpr\linewidth-2\fboxsep\relax}{\raggedright same-database isolation between queue instances}} &
\nolinkurl{_concurrent_success_case}; \nolinkurl{_instance_lock_failure_case};
\nolinkurl{_cross_instance_failure_case}; \nolinkurl{_concurrency_case} \\
\addlinespace
\colorbox{tracebaseline}{\parbox{\dimexpr\linewidth-2\fboxsep\relax}{\raggedright safe concurrent claimers}} &
\nolinkurl{_concurrent_success_case}; \nolinkurl{_cross_instance_failure_case};
\nolinkurl{_concurrency_case}; relevant upstream multithreading tests \\
\addlinespace
\colorbox{tracebaseline}{\parbox{\dimexpr\linewidth-2\fboxsep\relax}{\raggedright compatibility with sibling queue types}} &
\nolinkurl{_unique_ack_deduplication_case}; \nolinkurl{_sqlite_update_persistence_case};
\nolinkurl{_sqlite_delete_reopen_case}; \nolinkurl{_filo_sqlite_order_case};
\nolinkurl{_unique_sqlite_deduplication_case}; 89 upstream regression
coordinates \\
\addlinespace
\colorbox{tracebaseline}{\parbox{\dimexpr\linewidth-2\fboxsep\relax}{\raggedright single-item \texttt{get}, \texttt{ack},
\texttt{nack}, and \texttt{ack\_failed} operations}} &
\nolinkurl{_ack_state_regression_case}; preexisting single \texttt{get} in
\nolinkurl{_bulk_identity_case}; upstream ack/get/nack tests \\
\addlinespace
\colorbox{tracebaseline}{\parbox{\dimexpr\linewidth-2\fboxsep\relax}{\raggedright each instance's \texttt{qsize()} under the baseline
contract.}} & \nolinkurl{_success_case}; \nolinkurl{_bulk_identity_case}; all
concurrency functions; \nolinkurl{_ack_state_regression_case};
\nolinkurl{_sqlite_delete_reopen_case}; upstream size tests \\
\end{longtable}
\captionof{table}{PersistQueue prompt-to-test traceability.}
\label{tab:persistqueue-prompt-tests}
}
\section{Executable Diagnostic Catalog}
\label{app:diagnostic-catalogs}

These are the diagnostic checks that enter the four-axis summaries. The Oracle
alone determines overall correctness; each coordinate below executes one
targeted behavior and has no verdict authority. We omit the duplicated listing
of ordinary upstream regression-test names: those tests remain in the released
machine-readable catalog, while Table~\ref{tab:coordinate-audit} records their
measurement-audit disposition.

\begin{center}
\small
\begin{tabularx}{\linewidth}{@{}>{\raggedright\arraybackslash}p{0.17\linewidth}>{\raggedright\arraybackslash}p{0.30\linewidth}X@{}}
\toprule
\textbf{Task} & \textbf{Axis} & \textbf{Selected executable coordinates}\\
\midrule
\multirow{4}{*}{PersistQueue} & Recovery & Replacement compensation under FIFO and FILO.\\
& Identity & Successful raw/non-raw claims under FIFO/FILO; bulk identity; batch contract; unique-queue acknowledgement.\\
& Isolation & Cross-instance failure under FIFO/FILO; concurrent failure under FILO.\\
& Durability & Ack-state regression; SQLite update and delete/reopen; FILO order; unique SQLite deduplication.\\
\midrule
\multirow{4}{*}{Dijkstra} & Routing & Single-source reference; shared-junction distances, paths, source order, and target; source switching; competing routes; two-junction and false-success target/all-target checks; late-edge denial.\\
& Weights & Graph/multigraph shapes; parallel edges; missing/callable weights; zero weights/cycles; cutoff boundaries; filtered and unfiltered negative-weight behavior.\\
& API & Source identity and graph immutability; target-is-source; validation errors; callback exception identity.\\
& Compatibility & Medium branching; NetworkX 3.6 weighted paths, Voronoi cells, group closeness, and Steiner tree without a filter.\\
\bottomrule
\end{tabularx}
\captionof{table}{Selected coordinates grouped by diagnostic axis.}
\label{tab:coordinate-catalog}
\end{center}

The complete function names, source hashes, and one-sentence execution
descriptions are available in the public repository at
\href{https://poisonvalleys.org}{\textbf{poisonvalleys.org}}.
\section{Design Coverage and Robustness Detail}
\label{app:visual-placeholders}
\subsection{Outcomes by model configuration}

The table below is descriptive: conditions and poison-family mixtures differ
across rows, so it must not be read as a general model ranking.

\begin{center}
\small
\setlength{\tabcolsep}{4pt}
\begin{tabular}{@{}llrrr@{}}
\toprule
\textbf{Task} & \textbf{Requested model} & \textbf{Candidates} & \textbf{Success} & \textbf{Failure}\\
\midrule
\multirow{3}{*}{PersistQueue}
 & Claude Haiku 4.5 & 174 & 2 (\textcolor{green!50!black}{\textbf{1.1\%}}) & 172 (\textcolor{red!70!black}{\textbf{98.9\%}})\\
 & GPT-5.4 Nano, medium & 322 & 216 (\textcolor{green!50!black}{\textbf{67.1\%}}) & 106 (\textcolor{red!70!black}{\textbf{32.9\%}})\\
 & GPT-5.6 Luna, medium & 339 & 235 (\textcolor{green!50!black}{\textbf{69.3\%}}) & 104 (\textcolor{red!70!black}{\textbf{30.7\%}})\\
\midrule
\multirow{3}{*}{Dijkstra}
 & Claude Sonnet 5, medium & 331 & 91 (\textcolor{green!50!black}{\textbf{27.5\%}}) & 240 (\textcolor{red!70!black}{\textbf{72.5\%}})\\
 & GPT-5.6 Luna, medium & 340 & 119 (\textcolor{green!50!black}{\textbf{35.0\%}}) & 221 (\textcolor{red!70!black}{\textbf{65.0\%}})\\
 & Qwen 3.8 Flash & 127 & 6 (\textcolor{green!50!black}{\textbf{4.7\%}}) & 121 (\textcolor{red!70!black}{\textbf{95.3\%}})\\
\bottomrule
\end{tabular}
\captionof{table}{Candidate outcomes by requested model configuration.}
\label{tab:model-outcomes}
\end{center}

\subsection{Response availability by stage and requested model}
The UUID ledger, not the filesystem inventory, defines every scientific
denominator. Its global funnel is 1,844 planned = 1,633 code-bearing/execution-
reaching + 211 no candidate. The latter divides into 203 response-unavailable
rows and eight response-available rows with no executable candidate; 1,340 of
the 1,633 completed the authoritative Oracle. Separately, the frozen run trees
contain 1,678 response files and 1,815 \texttt{attempt.json} records. Those
are file inventories, not alternate populations and not terms in the
attrition identities below.

\begin{center}
\small
\begin{tabular}{@{}lllrrr@{}}
\toprule
Task & Stage & Requested model & Planned & Code & No candidate \\
\midrule
Dijkstra & canonical & not recorded & 64 & 0 & 64 \\
Dijkstra & canonical & Claude Sonnet 5 & 9 & 9 & 0 \\
Dijkstra & canonical & GPT-5.6 Luna & 10 & 10 & 0 \\
Dijkstra & canonical & Qwen 3.8 Flash & 2 & 2 & 0 \\
Dijkstra & H-only & Claude Sonnet 5 & 80 & 80 & 0 \\
Dijkstra & H-only & GPT-5.6 Luna & 80 & 80 & 0 \\
Dijkstra & H-only & Qwen 3.8 Flash & 29 & 29 & 0 \\
Dijkstra & P-only & Claude Sonnet 5 & 76 & 76 & 0 \\
Dijkstra & P-only & GPT-5.6 Luna & 80 & 80 & 0 \\
Dijkstra & P-only & Qwen 3.8 Flash & 32 & 32 & 0 \\
Dijkstra & H+P & Claude Sonnet 5 & 79 & 79 & 0 \\
Dijkstra & H+P & GPT-5.6 Luna & 80 & 80 & 0 \\
Dijkstra & H+P & Qwen 3.8 Flash & 28 & 28 & 0 \\
Dijkstra & no-plan & Claude Sonnet 5 & 87 & 87 & 0 \\
Dijkstra & no-plan & GPT-5.6 Luna & 90 & 90 & 0 \\
Dijkstra & no-plan & Qwen 3.8 Flash & 36 & 36 & 0 \\
PersistQueue & canonical & Claude Haiku 4.5 & 4 & 4 & 0 \\
PersistQueue & canonical & GPT-5.4 Nano & 11 & 10 & 1 \\
PersistQueue & canonical & GPT-5.6 Luna & 10 & 10 & 0 \\
PersistQueue & H-only & Claude Haiku 4.5 & 55 & 52 & 3 \\
PersistQueue & H-only & GPT-5.4 Nano & 118 & 108 & 10 \\
PersistQueue & H-only & GPT-5.6 Luna & 111 & 110 & 1 \\
PersistQueue & P-only & Claude Haiku 4.5 & 55 & 54 & 1 \\
PersistQueue & P-only & GPT-5.4 Nano & 116 & 101 & 15 \\
PersistQueue & P-only & GPT-5.6 Luna & 111 & 110 & 1 \\
PersistQueue & H+P & Claude Haiku 4.5 & 66 & 64 & 2 \\
PersistQueue & H+P & GPT-5.4 Nano & 164 & 103 & 61 \\
PersistQueue & H+P & GPT-5.6 Luna & 161 & 109 & 52 \\
\bottomrule
\end{tabular}
\end{center}
PersistQueue source absence is concentrated in the H+P cells for GPT-5.4 Nano and GPT-5.6 Luna and is classified by provider-specific terminal subcategories. Dijkstra's 64 no-candidate rows are legacy canonical rows without requested-model metadata; all designed poison-stage cells are code-bearing.
\subsection{Leave-one-group-out sensitivity}
\begin{center}
\small
\begin{tabular}{@{}llrr@{}}
\toprule
Task & Omission & P-only $-$ H-only RD range & H+P $-$ P-only RD range \\
\midrule
PersistQueue & Omit one family & +28.8 to +36.7 pp & +5.8 to +9.3 pp \\
PersistQueue & Omit one model & +24.6 to +40.9 pp & +1.9 to +13.2 pp \\
Dijkstra & Omit one family & +26.9 to +31.8 pp & -1.5 to +3.7 pp \\
Dijkstra & Omit one model & +22.5 to +33.2 pp & +0.8 to +2.3 pp \\
\bottomrule
\end{tabular}
\end{center}
The primary P-only contrast remains positive under every leave-one-family and leave-one-model omission in both tasks. The smaller H+P increment is less stable, especially in Dijkstra.
\subsection{Diagnostic coverage and fidelity boundary}
\begin{center}
\small
\begin{tabular}{@{}lrrr@{}}
\toprule
Task & Complete diagnostic panel & Incomplete panel & No code \\
\midrule
PersistQueue & 660 & 175 & 147 \\
Dijkstra & 669 & 129 & 64 \\
\bottomrule
\end{tabular}
\end{center}
Diagnostic incompleteness is an evidence state, not a terminal failure class. Every terminal failure still has a specific cause subcategory. The mechanical source screen covers assessable code but has zero adjudicated rows, so literal-realization rates are not estimable and no fidelity bar chart is shown.
\section{Statistical, Measurement, and Reproducibility Detail}
\label{app:statistical-placeholders}
\subsection{Frozen inputs and regeneration contract}
The binary ledger manifest is \texttt{f5dba9161a5fc4d7\ldots}; the corrected raw-coordinate manifest is \texttt{1b9b7b8a3c0e0e4c\ldots}. The regeneration script verifies those manifests and the exact attempt CSV hashes, reconciles 1,844 UUIDs, rejects any terminal FAIL without a specific subcategory, and writes all tables below deterministically. Family--phenotype intervals use 2,000 attempt-level bootstrap samples with seed 20260903. For stage $s$ and common-support cell $g$, the balanced failure rate is $\hat p_s=\sum_g w_g\hat p_{s,g}$ with the same cell weights $w_g$ in both stages; we report $\hat p_P-\hat p_H$ and $\hat p_P/\hat p_H$. Primary intervals come from a dependence-group cluster bootstrap that resamples complete request UUIDs.
\subsection{Provider and execution settings}
\begin{center}
\scriptsize
\begin{tabular}{@{}lrrrrl@{}}
\toprule
\textbf{Requested profile} & \textbf{Max tokens} & \textbf{Reasoning} & \textbf{Retries} & \textbf{Timeouts (request/exec.)} & \textbf{Settings SHA prefix} \\
\midrule
Claude Haiku 4.5 & 64000 & none & 0 & 600/600 s & \texttt{143903211a84} \\
Claude Sonnet 5 & 128000 & medium & 0 & 600/600 s & \texttt{0d8cb11824ef} \\
Claude Sonnet 5 & 128000 & medium & 0 & 600/600 s & \texttt{d76c4a15588a} \\
GPT-5.4 Nano & 128000 & medium & 0 & 600/600 s & \texttt{9459a5981a5e} \\
GPT-5.6 Luna & 128000 & medium & 0 & 600/600 s & \texttt{89090236fc0b} \\
Qwen 3.8 Flash & 131072 & unspecified & 0 & 600/600 s & \texttt{69c09d376571} \\
\bottomrule
\end{tabular}
\captionof{table}{Provider and execution settings.}
\label{tab:runtime-settings}
\end{center}

The final analysis ledger preserves both the requested API profile and the model
identifier returned by the provider. Table~\ref{tab:model-identifier-audit}
reports these fields separately: they are not Codex task or agent names. The
OpenAI Nano alias is returned as the dated snapshot
\texttt{gpt-5.4-nano-2026-03-17}. A missing returned identifier is recorded as
\texttt{NOT\_OBSERVED}, never imputed from the request.

\begin{center}
\scriptsize
\begin{tabular}{@{}llr@{}}
\toprule
\textbf{Identity field} & \textbf{Recorded value} & \textbf{Rows} \\
\midrule
Requested & \texttt{gpt-5.6-luna} & 733 \\
Requested & \texttt{gpt-5.4-nano} & 409 \\
Requested & \texttt{claude-sonnet-5} & 331 \\
Requested & \texttt{claude-haiku-4-5-20251001} & 180 \\
Requested & \texttt{qwen3.8-flash} & 127 \\
Requested & \texttt{NOT\_APPLICABLE} & 64 \\
\midrule
Returned & \texttt{gpt-5.6-luna} & 679 \\
Returned & \texttt{gpt-5.4-nano-2026-03-17} & 327 \\
Returned & \texttt{claude-sonnet-5} & 331 \\
Returned & \texttt{claude-haiku-4-5-20251001} & 177 \\
Returned & \texttt{qwen3.8-flash} & 127 \\
Returned & \texttt{NOT\_OBSERVED} & 203 \\
\bottomrule
\end{tabular}
\captionof{table}{Requested and provider-returned model identifiers.}
\label{tab:model-identifier-audit}
\end{center}

Requests were issued on August 31, 2026 through OpenAI, Anthropic, and Alibaba/Qwen provider
adapters. Provider retries were zero. Anthropic used
\texttt{standard\_only} with storage disabled, OpenAI used the default service
tier with storage disabled, and Qwen sent neither a service tier nor a storage
request. Temperature was not sent for the factorial profiles. The locked
environment specifies Python $\geq$3.13 and $<3.14$, Anthropic SDK 1.2.0,
OpenAI SDK 2.54.0, and HTTPX 0.28.1.

\begin{center}
\small
\begin{tabular}{@{}lr@{}}
\toprule
\textbf{Record type} & \textbf{Files} \\
\midrule
Provider requests & 1,844 \\
Provider responses & 1,678 \\
Attempt manifests & 1,815 \\
Execution receipts & 1,633 \\
Candidate-subprocess records & 1,633 \\
Attempt logs & 1,844 \\
Run manifests / summaries / finalization records & 3 / 3 / 3 \\
Unreadable selected structured artifacts & 0 \\
\bottomrule
\end{tabular}
\captionof{table}{Inventory of the four frozen run trees.}
\label{tab:frozen-log-inventory}
\end{center}

The final analysis ledger is authoritative for scientific denominators: its
1,844 rows are keyed by unique \texttt{attempt\_id}, and its explicit inclusion
flags reproduce 1,641 provider-available responses, 203 unavailable responses,
1,633 code-bearing and execution-reaching attempts, and 1,340 completed
authoritative-Oracle evaluations. The identities $1{,}844=1{,}641+203$ and
$1{,}641=1{,}633+8$ hold exactly. By contrast, the 1,678 response artifacts and
1,815 attempt records in Table~\ref{tab:frozen-log-inventory} are filesystem
inventories. Because the surviving response and attempt-record formats do not
consistently expose the paper's analytical \texttt{attempt\_id}, those inventory
counts are not silently converted into UUID-level populations.

The attempt manifests contain 1,815 observations of
\texttt{provider.request\_elapsed\_seconds}. Table~\ref{tab:request-latency-audit}
summarizes those preserved receipts. Their durations sum to approximately
218,629 seconds, or 60.73 aggregate provider-request hours. This sum is neither
experiment wall time nor CPU-hours: requests may overlap, and this receipt
population has not been substituted for the final 1,844-row analysis ledger.

\begin{center}
\small
\begin{tabular}{@{}lrrrr@{}}
\toprule
\textbf{Requested profile} & $n$ & \textbf{Min. (s)} & \textbf{Median (s)} & \textbf{Max. (s)} \\
\midrule
Claude Haiku 4.5 & 177 & 29.32 & 33.94 & 87.04 \\
Claude Sonnet 5 & 340 & 197.70 & 291.27 & 441.27 \\
GPT-5.4 Nano & 409 & 0.86 & 79.83 & 601.91 \\
GPT-5.6 Luna & 734 & 0.90 & 38.54 & 604.14 \\
Qwen 3.8 Flash & 155 & 49.62 & 237.28 & 482.03 \\
\bottomrule
\end{tabular}
\captionof{table}{Provider-request latency in preserved attempt receipts.}
\label{tab:request-latency-audit}
\end{center}
\subsection{Measurement audit}
\begin{center}
\small
\begin{tabular}{@{}lrrrr@{}}
\toprule
\textbf{Task} & \textbf{Validated} & \textbf{With limitations} & \textbf{Insufficient} & \textbf{Corrected assignment} \\
\midrule
PersistQueue & 0 & 18 & 97 & 0 \\
Dijkstra & 10 & 12 & 5 & 1 \\
\bottomrule
\end{tabular}
\captionof{table}{Coordinate construct-audit disposition.}
\label{tab:coordinate-audit}
\end{center}

The row and reducer pipeline is deterministic and UUID-complete, but construct validity is not uniformly established. PersistQueue's 18 historical reducer members are all \texttt{VALID\_WITH\_LIMITATIONS}; Dijkstra has ten validated coordinates, twelve validated with limitations, five with insufficient evidence, and one corrected routing-origin assignment. The source-pattern screen supplies a sampling frame, but its adjudicated count is 0; therefore no adoption or rejection rate is reported.
The authority boundary is fixed: selected Oracles determine binary verdicts;
only stated reducer coordinates determine complete phenotypes; descriptive
extensions cannot change either result. The current Dijkstra reducer includes
the versioned corrected routing-origin assignment. PersistQueue reduction stays
within the selected historical panel, and its descriptive extensions remain
outside that boundary.
\subsection{Specific terminal failure subcategories}
\begingroup\scriptsize
\begin{longtable}{@{}p{0.13\textwidth}p{0.28\textwidth}p{0.43\textwidth}r@{}}
\toprule Task & Primary category & Specific subcategory & Count \\
\midrule\endfirsthead
\toprule Task & Primary category & Specific subcategory & Count \\
\midrule\endhead
\bottomrule\endfoot
Dijkstra & NO\_CANDIDATE\_ARTIFACT & \nolinkurl{MISSING_CANDIDATE_SOURCE} & 64 \\
Dijkstra & ORACLE\_ASSERTION\_FAILURE & \nolinkurl{false_success_target_projection} & 1 \\
Dijkstra & ORACLE\_ASSERTION\_FAILURE & \nolinkurl{medium_branching_projection} & 1 \\
Dijkstra & ORACLE\_ASSERTION\_FAILURE & \nolinkurl{networkx_3_6_error_regression} & 166 \\
Dijkstra & ORACLE\_ASSERTION\_FAILURE & \nolinkurl{shared_junction_combined_distances} & 264 \\
Dijkstra & ORACLE\_ASSERTION\_FAILURE & \nolinkurl{shared_junction_combined_paths} & 1 \\
Dijkstra & ORACLE\_ASSERTION\_FAILURE & \nolinkurl{single_source_reference_route} & 11 \\
Dijkstra & ORACLE\_ASSERTION\_FAILURE & \nolinkurl{two_junction_all_target_projection} & 8 \\
Dijkstra & ORACLE\_EXECUTION\_FAILURE & \nolinkurl{SUBPROCESS_CRASH:AttributeError} & 17 \\
Dijkstra & ORACLE\_EXECUTION\_FAILURE & \nolinkurl{SUBPROCESS_CRASH:ImportError} & 16 \\
Dijkstra & ORACLE\_EXECUTION\_FAILURE & \nolinkurl{SUBPROCESS_CRASH:KeyError} & 57 \\
Dijkstra & ORACLE\_EXECUTION\_FAILURE & \nolinkurl{SUBPROCESS_CRASH:NameError} & 9 \\
Dijkstra & ORACLE\_EXECUTION\_FAILURE & \nolinkurl{SUBPROCESS_CRASH:RecursionError} & 4 \\
Dijkstra & ORACLE\_EXECUTION\_FAILURE & \nolinkurl{SUBPROCESS_CRASH:TypeError} & 12 \\
Dijkstra & ORACLE\_EXECUTION\_FAILURE & \nolinkurl{SUBPROCESS_CRASH:UNAVAILABLE} & 1 \\
Dijkstra & ORACLE\_EXECUTION\_FAILURE & \nolinkurl{SUBPROCESS_CRASH:ValueError} & 2 \\
Dijkstra & ORACLE\_EXECUTION\_FAILURE & \nolinkurl{SUBPROCESS_CRASH:networkx.exception.NetworkXNoPath} & 9 \\
Dijkstra & ORACLE\_EXECUTION\_FAILURE & \nolinkurl{TIMEOUT_600_SECONDS} & 3 \\
PersistQueue & CANDIDATE\_RUNTIME\_EXCEPTION & \nolinkurl{CANDIDATE_RUNTIME_EXCEPTION} & 29 \\
PersistQueue & EXECUTION\_TIMEOUT & \nolinkurl{PUBLIC_ORACLE_SUBPROCESS_TIMEOUT} & 1 \\
PersistQueue & NO\_CANDIDATE\_ARTIFACT & \nolinkurl{PROVIDER_CONNECTION_ERROR} & 2 \\
PersistQueue & NO\_CANDIDATE\_ARTIFACT & \nolinkurl{PROVIDER_RATE_LIMIT_CREDIT_EXHAUSTED} & 105 \\
PersistQueue & NO\_CANDIDATE\_ARTIFACT & \nolinkurl{PROVIDER_REQUEST_TIMEOUT} & 32 \\
PersistQueue & ORACLE\_CHECK\_FAILURE & \nolinkurl{same_invocation::custom::replacement_compensation::fifo} & 215 \\
PersistQueue & ORACLE\_CHECK\_FAILURE & \nolinkurl{same_invocation::custom::replacement_compensation::filo} & 4 \\
PersistQueue & ORACLE\_EXECUTION\_EXCEPTION & \nolinkurl{PUBLIC_ORACLE_EXCEPTION} & 133 \\
PersistQueue & RECONSTRUCTION\_FAILURE & \nolinkurl{RECON_INVALID_CODE} & 7 \\
PersistQueue & RECONSTRUCTION\_FAILURE & \nolinkurl{RECON_SENTINEL_MISMATCH} & 1 \\
\end{longtable}
\captionof{table}{Exhaustive terminal failure taxonomy.}
\label{tab:failure-subcategories}
\endgroup
\subsection{Complete design cells}
The immutable ledger contains every nonempty family $\times$ stage $\times$
dependency $\times$ requested-model cell. Printing hundreds of ledger rows here
duplicates the released data without helping a reader compare conditions.
The exact table, including zero-cell complements and planned/code/failure
counts, is available as \texttt{full-design-cells.csv} in the public repository:
\href{https://poisonvalleys.org}{\textbf{poisonvalleys.org}}.
The stage-by-model coverage table in Appendix~\ref{app:visual-placeholders}
summarizes the imbalance that motivates common-support standardization.
\end{document}